\documentclass[11pt]{article}
\usepackage{amssymb,cite}
\newcommand{\be}{\begin{eqnarray}}
\newcommand{\ee}{\end{eqnarray}}
\newcommand{\bez}{\begin{eqnarray*}}
\newcommand{\eez}{\end{eqnarray*}}
\newcommand{\pa}{\partial}
\newcommand{\la}{\lambda}

\renewcommand{\theequation} {\arabic{section}.\arabic{equation}}

\title{\bf Explorations of the Extended ncKP Hierarchy}

\date{  }

\author{Aristophanes Dimakis \\
 Department of Financial and Management Engineering, \\
 University of the Aegean, 31 Fostini Str., GR-82100 Chios, Greece \\
 dimakis@aegean.gr
          \and
 Folkert M\"uller-Hoissen \\ Max-Planck-Institut f\"ur Str\"omungsforschung \\
 Bunsenstrasse 10, D-37073 G\"ottingen, Germany \\
 fmuelle@gwdg.de }

\begin{document}

\newtheorem{theorem}{Theorem}[section]
\newtheorem{lemma}{Lemma}[section]

\maketitle

\begin{abstract}
A recently obtained extension (xncKP) of the Moyal-deformed KP hierarchy (ncKP hierarchy)
by a set of evolution equations in the Moyal-deformation parameters is further explored.
Formulae are derived to compute these equations efficiently.
Reductions of the xncKP hierarchy are treated, in particular to the extended
ncKdV and ncBoussinesq hierarchies. Furthermore, a good part of the Sato formalism
for the KP hierarchy is carried over to the generalized framework.
In particular, the well-known bilinear identity theorem for the KP hierarchy,
expressed in terms of the (formal) Baker-Akhiezer function, extends to the xncKP hierarchy.
Moreover, it is demonstrated that $N$-soliton solutions of the ncKP equation are
also solutions of the first few deformation equations. This is shown to be related
to the existence of certain families of algebraic identities.
\end{abstract}

\small
\tableofcontents
\normalsize

\section{Introduction}
\setcounter{equation}{0}
The \emph{noncommutative KP hierarchy} (see \cite{Kupe00}, in particular)
is defined as the set of equations
\be
    L_{t_n} := \frac{\partial L}{\partial t_n}
             = L^{(n)} \ast L - L \ast L^{(n)}
             =: [L^{(n)},L]_\ast = [ L , \bar{L}^{(n)}]_\ast
             \qquad   n = 1,2, \ldots  \label{ncKPh}
\ee
in terms of formal pseudo-differential operators
\be
   L = \pa + \sum_{k=1}^\infty u_{k+1} \, \pa^{-k} \qquad
   L^{(n)} = (L^n)_{\geq 0}  \qquad
   \bar{L}^{(n)} = (L^n)_{<0} = L^n-L^{(n)}   \label{L-u}
\ee
with (matrices of) functions $u_k$ and $L^n = L^{n-1} \ast L$.
The non-negative (negative) part of a formal series
is understood in the sense of non-negative (negative) powers of the operator $\pa$
of partial differentiation with respect to $x = t_1$. $\pa$ has to be a
derivation of the associative product $\ast$. In this work, we assume that
the product is given by
\be
   f \ast g = \mathbf{m} \circ e^{P/2} (f \otimes g)
  \qquad
   P = \sum_{m,n=1}^\infty \theta_{m,n} \, \pa_{t_m} \otimes \pa_{t_n}
   \label{Moyal}
\ee
where $\mathbf{m}(f \otimes g) = f \, g$ for functions $f,g$, and
$\theta_{n,m} = - \theta_{m,n}$ are constants. This implies
\be
   (f \ast g)_{\theta_{m,n}} = f_{\theta_{m,n}} \ast g + f \ast g_{\theta_{m,n}}
   + {1 \over 2} ( f_{t_m} \ast g_{t_n} - f_{t_n} \ast g_{t_m} ) \; .
    \label{dtheta_star}
\ee
The \emph{ncKP hierarchy} obtained in this way \cite{Hama03b,DMH04hier} is thus
a Moyal-deformation of the classical KP hierarchy (see
\cite{Sato+Sato82,DKJM83,Sega+Wils85,BBT03,Dick03,OSTT88}, for example).
Such deformations of soliton equations have recently been discussed in several papers
(see \cite{Hama03b,DMH04hier,SWW04,LMPPT04} and the references cited there), partly
motivated by the appearance of related structures in string theory.
\vskip.1cm

It has been shown in \cite{DMH04hier} that the ncKP hierarchy can be extended to a
bigger hierarchy, called \emph{xncKP hierarchy} in the following, by including the further
(deformation) equations
\be
    L_{\theta_{m,n}}
 = [W^{(m,n)},L]_\ast + {1 \over 2} \Big( L_{t_n} \ast L^{(m)} - L_{t_m} \ast L^{(n)} \Big)
   \label{L_theta}
\ee
where
\be
    W^{(m,n)}
  = {1 \over 2} \Big( \bar{L}^{(n)} \ast L^{(m)} - \bar{L}^{(m)} \ast L^{(n)}\Big)_{\geq 0} \; .
    \label{Wmn}
\ee
The corresponding flows commute with those of the ncKP hierarchy. More precisely,
for fixed natural numbers $k,m,n$, the $\theta_{m,n}$-flow commutes with the
$t_k$-flow if (\ref{ncKPh}) holds for $k,m,n$.
Furthermore, the deformation flows also commute with each other if the associated ncKP
hierarchy equations hold. To be more precise, if we fix four natural numbers $k,l,m,n$,
the $\theta_{k,l}$-flow commutes with the $\theta_{m,n}$-flow if (\ref{ncKPh}) holds
for $k,l,m,n$. The extension of the ncKP hierarchy considered here may
therefore be regarded as being of `second order'. It is \emph{not} a direct extension
of the ncKP hierarchy in the sense in which each of its members extends the set of
remaining equations.
\vskip.1cm

The xncKP hierarchy equations are the integrability conditions of the linear system
\be
    L \ast \psi = \lambda \, \psi   \qquad
    \psi_{t_n} = L^{(n)} \ast \psi  \qquad
    \psi_{\theta_{m,n}} = W^{(m,n)} \ast \psi  \; .    \label{xncKP_ls}
\ee
In the case of the `commutative' KP hierarchy, a solution $\psi$ of the first two
of equations (\ref{xncKP_ls}) is given by the so-called (formal) \emph{Baker-Akhiezer function}
of the KP hierarchy, which plays a crucial role in the Sato formalism
\cite{DKJM83,Sega+Wils85,BBT03,Dick03,OSTT88}.
In section~\ref{sec:B-A} we introduce a Baker-Akhiezer function for the ncKP hierarchy
which will be important in the subsequent sections.
\vskip.1cm

The construction of the Baker-Akhiezer function involves a pseudo-differential operator
with which $L$ is written as a dressing of $\pa$. An important step in Sato theory is to express
the hierarchy equations in terms of this operator
\cite{Sato+Sato82,DKJM83,Sega+Wils85,BBT03,Dick03,OSTT88}.
In section~\ref{sec:af} we derive a corresponding formulation of the xncKP hierarchy.
Section~\ref{sec:bil} then proves, in particular, that the xncKP hierarchy equations
can be cast into \emph{bilinear identities}, which again generalizes a classical result
(see \cite{BBT03,Dick03}, for example).
\vskip.1cm

Equations (\ref{ncKPh}) determine the $u_k$, $k>2$, in terms of $u_2$
(see appendix A). Introducing a potential $\phi$ via
\be
    u_2 = \phi_x      \label{u2pot}
\ee
and taking the residue\footnote{The residue of a formal series is the coefficient
of the $\pa^{-1}$ term in the series. Note that it is irrelevant on which side of
$\pa^{-1}$ one reads off the coefficient.}
of (\ref{ncKPh}), results in a set of evolution equations for $\phi$:\footnote{Here
and in the following, possible constants of $x$-integration are set to zero.
This can be substantiated with the assumption that the fields and
their derivatives vanish as $|x| \to \infty$. Note that the first ($n=1$) of equations
(\ref{ncKPeqs}) is an identity.}
\be
   \phi_{t_n} = \mbox{res} (L^n)  \qquad n = 1,2, \ldots \; .   \label{ncKPeqs}
\ee
Expressions for the first few residues appearing on the right hand side
in terms of the $u_k$ are given in appendix B. In section~\ref{sec:recursion} we derive
more convenient expressions for the above equations and, more generally,
for those of the xncKP hierarchy. In particular, via the detour through the extension
of the ncKP hierarchy, we find formulae for the ncKP equations, which reduce their
computation to certain recursion relations. These results no longer refer to the
extension and are actually not restricted to the special choice of
product (\ref{Moyal}). Thus, Moyal-deformation and extension in the aforementioned sense
may even lead to new insights into the classical hierarchies.
\vskip.1cm

Section~\ref{sec:xncKP} presents some concrete examples of xncKP hierarchy equations.
Here we concentrate on expressing them in the form $\phi_{t_n} = K_n$,
$\phi_{\theta_{m,n}} = K_{m,n}$, where the right hand sides are expressed solely
in terms of the potential $\phi$ and its derivatives with respect
to $x$ and $y=t_2$, at the expense of having to allow for $x$-integrals also.
\vskip.1cm

Applying reduction methods to integrable equations leads to other integrable equations.
Section~\ref{sec:red} tackles the question of what we obtain in this way from the new
(deformation) equations. In particular, we consider the reduction of the xncKP hierarchy
to the extended ncKdV and ncBoussinesq hierarchies.
\vskip.1cm

Given any solution of the KP equation (or another member of its hierarchy), a deformation
equation allows us to compute a corresponding (formal) solution of the ncKP equation which
reduces to the former at vanishing deformation parameter. This is done by calculating
iteratively higher derivatives of $\phi$ with respect to $\theta_{m,n}$ at $\theta_{m,n} = 0$
and writing down a formal Taylor series. Because of the commutativity of the flows, this yields
indeed a (formal) solution of the ncKP equation for any given initial KP data.\footnote{See
also \cite{DMH00ncKdV} for a corresponding calculation in case of the ncKdV equation.}
In such an approach, there is hardly a chance to solve the corresponding equations to
all orders in the respective deformation parameter. However, using a power series expansion
in a new parameter $\epsilon$, deformed soliton solutions of the ncKP equation were indeed
obtained to all orders in $\epsilon$ \cite{Pani01}. In section~\ref{sec:ncKPsol} we extend
this method to some of the deformation equations. We also refer to
\cite{Wang+Wada03dbar,Wang+Wada03ncKP,Wang+Wada03ncsoliton} for other methods and
special solutions of the ncKP equation.
\vskip.1cm

Finally, section~\ref{sec:concl} contains some concluding remarks.
Some supplementary material has been separated from the main text as a series
of appendices (A-F), in order to achieve a better readability of the main text.

\section{Baker-Akhiezer function and its adjoint}
\label{sec:B-A}
\setcounter{equation}{0}
Let us write the Lax operator $L$ as a `dressing' of $\pa$:
\be
    L = X \ast \pa \ast X^{-1}     \label{LX}
\ee
where $X$ is an invertible (formal) pseudo-differential operator
\be
    X = 1 + \sum_{i=1}^\infty w_i(t,\theta) \, \pa^{-i}
\ee
with (matrices of) functions $w_i$.\footnote{See appendix C for formulae
expressing the $w_i$ in terms of the coefficients $u_k$ of $L$.}
An important step in the Sato formalism is to express the KP hierarchy in
terms of $X$ (see \cite{BBT03,Dick03}, for example).
We prove this result for the ncKP hierarchy.

\begin{theorem}
\label{theorem:XncKPh}
The ncKP hierarchy equations (\ref{ncKPh}) are equivalent to
\be
   X_{t_n}  = - (X \ast \pa^n \ast X^{-1})_{<0} \ast X  \; . \label{XncKPh}
\ee
\end{theorem}
{\em Proof:} By differentiation of (\ref{LX}), we obtain the identity
$L_{t_n} = [ X_{t_n} \ast X^{-1} , L ]_\ast$. If (\ref{XncKPh}) holds, using
$L^n = X \ast \pa^n \ast X^{-1}$ it follows that (\ref{ncKPh}) is satisfied.
Let us now assume that (\ref{ncKPh}) holds. Inserting (\ref{LX})
in (\ref{ncKPh}), after some manipulations we obtain
\bez
   [ X^{-1} \ast ( X_{t_n} + (X \ast \pa^n \ast X^{-1})_{<0} \ast X ) ]_x = 0
\eez
which implies
\bez
   X_{t_n} + (X \ast \pa^n \ast X^{-1})_{<0} \ast X = X \ast C_n
\eez
with $C_n = \sum_{i=1}^\infty C_{n,i} \, \pa^{-i}$ independent of $x$.
Differentiation with respect to $t_m$ and using
$(\bar{L}^{(m)})_{t_n} - (\bar{L}^{(n)})_{t_m} = [\bar{L}^{(m)},\bar{L}^{(n)}]_\ast$,
which follows from (\ref{ncKPh}) (see also \cite{DMH04hier}), leads to
$(C_n)_{t_m} - (C_m)_{t_n} + [C_m,C_n]_\ast = 0$ and thus $C_n = C^{-1} \ast C_{t_n}$
with $C = 1 + \sum_{i=1}^\infty c_i \, \pa^{-i}$ independent of $x$. Hence
\bez
    X_{t_n} + (X \ast \pa^n \ast X^{-1})_{<0} \ast X = X \ast C^{-1} \ast C_{t_n}
\eez
and a transformation $X \to X \ast C$ yields (\ref{XncKPh}).\footnote{$X$ is determined by $L$
via (\ref{LX}) up to transformations $X \to X \ast C$, where $C = 1 + \sum_{i=1}^\infty c_i \, \pa^{-i}$
with $c_i$ independent of $x$. For the commutative case a proof can be found in \cite{BBT03},
p. 342. This is easily adapted to the noncommutative case under consideration.}
\hfill $\blacksquare$
\vskip.2cm

Let $\xi = \sum_{i=1}^\infty t_i \, \la^i$. Since this is linear in the variables $t_i$,
we have $f(\xi) \ast g(\xi) = f(\xi) \, g(\xi)$. In particular, $e^\xi \ast e^{-\xi} = 1$.
Furthermore,
\be
   (e^\xi)_{t_n} = \la^n \, e^\xi   \qquad
    \pa^j e^\xi = \la^j \, e^\xi \qquad \forall j \in \mathbb{Z}
\ee
since $t_1 = x$. We define the \emph{Baker-Akhiezer function}\footnote{Here the operator $X$ is
evaluated on $e^\xi$. The latter is thus not treated as an operator.}
\be
 \psi = X \ast e^\xi = (1 + \sum_{i=1}^\infty w_i(t,\theta) \, \la^{-i}) \ast e^\xi
 = \hat{w}(t,\theta,\la) \ast e^\xi  \; .   \label{BAfunction}
\ee
It follows that $\psi$ satisfies the equations of the linear system of the ncKP hierarchy:
\be
 L \ast \psi &=& X \ast \pa \ast X^{-1} \ast X \ast e^\xi = \la \, X \ast e^\xi = \la \, \psi \\
 \psi_{t_n} &=& X_{t_n} \ast e^\xi + X \ast \la^n e^\xi
  = -\bar{L}^{(n)} \ast \psi + L^n \ast \psi = L^{(n)} \ast \psi
\ee
where we made use of (\ref{XncKPh}) in the form $X_{t_n} = - \bar{L}^{(n)} \ast X$.
\vskip.1cm

Next we introduce an \emph{involution}. For a function $f$, let
\be
    f(\theta)^\dag = f(-\theta)
\ee
(suppressing unaffected further arguments). As a consequence,
\be
   (f \ast g)^\dag = g^\dag \ast f^\dag
\ee
for any two functions $f,g$. This extends to pseudo-differential operators
via\footnote{The case of matrix coefficients is covered if ${}^\dag$ acts
as above on all matrix components, and also takes the transpose of the respective matrix.}
\be
   (f \pa^j)^\dagger = (-\pa)^j \, f^\dagger   \qquad \forall j \in \mathbb{Z}
\ee
as an involution:
\be
        (A \ast B)^\dag = B^\dag \ast A^\dag \; .
\ee
In particular, the adjoint of $L$ is given by
\be
   L^\dag = -(X^{-1})^\dag \ast \pa \ast X^\dag \; .
\ee
Furthermore, $1 = (X \ast X^{-1})^\dag = (X^{-1})^\dag \ast X^\dag$ and thus
$(X^{-1})^\dag = (X^\dag)^{-1}$. Let us also define an \emph{adjoint} of
the Baker-Akhiezer function:
\be
 \psi^\ast = (X^\dag)^{-1} \ast e^{-\xi}
           = \hat{w}^\ast(t,\theta,\la)^\dag \ast e^{-\xi}
             \label{BA^ast}
\ee
where we wrote
\be
  X^{-1} = 1 + \sum_{n=1}^\infty \pa^{-n} \, w^{(\ast)}_n  \qquad \quad
  \hat{w}^\ast = 1 + \sum_{n=1}^\infty w^{(\ast)}_n \, \lambda^{-n}   \label{hatw^ast}
\ee
with functions $w^{(\ast)}_n$. Using $(X^{-1})_{t_m} = X^{-1} \ast \bar{L}^{(n)}$,
it follows that
\be
  L^\dag \ast \psi^\ast = \la \, \psi^\ast   \qquad
  \psi^\ast_{t_n} = - L^{(n)}{}^\dag \ast \psi^\ast \; .  \label{ncKPls_ast}
\ee
It is often helpful to convert more generally (formal) pseudo-differential operators
into (formal) series in the variable $\la$, as done in (\ref{BAfunction})
and (\ref{BA^ast}):
\be
    A \ast e^{\la \, x} = \sum_j a_j \ast  \pa^j e^{\la \, x}
    = \sum_j \la^j \, a_j \ast e^{\la \, x}
\ee
where $A = \sum_j a_j \, \pa^j$.

\section{An alternative formulation of the xncKP hierarchy}
\label{sec:af}
\setcounter{equation}{0}
In section~\ref{sec:B-A} we derived an equivalent formulation
of the ncKP hierarchy equations (\ref{ncKPh}) in terms of the dressing operator $X$.
The following result shows that a corresponding formulation also exists for the
xncKP hierarchy.

\begin{theorem}
If (\ref{ncKPh}) (or equivalently (\ref{XncKPh}), see Theorem~\ref{theorem:XncKPh}) holds,
then (\ref{L_theta}) is equivalent to
\be
      X_{\theta_{m,n}}
   = - {1 \over 2} ( X_{t_m} \ast \pa^n \ast X^{-1} - X_{t_n} \ast \pa^m \ast X^{-1})_{<0} \ast X
     \; .  \label{dthetaX}
\ee
This in turn is then equivalent to
\be
      X_{\theta_{m,n}}
  &=& {1 \over 2} ( \bar{L}^{(m)} \ast L^n - \bar{L}^{(n)} \ast L^m )_{<0} \ast X \; .
      \label{dthetaXL}
\ee
\end{theorem}
{\em Proof:} Differentiation of $X \ast X^{-1}=1$ with respect to $\theta_{m,n}$, using
(\ref{dtheta_star}), leads to
\bez
  X_{\theta_{m,n}} \ast X^{-1} + X \ast (X^{-1})_{\theta_{m,n}}
  + {1 \over 2} (X_{t_m} \ast (X^{-1})_{t_n} - X_{t_n} \ast (X^{-1})_{t_m}) = 0
\eez
and thus, using (\ref{XncKPh}),
\bez
 X \ast (X^{-1})_{\theta_{m,n}} = - X_{\theta_{m,n}} \ast X^{-1}
  - {1 \over 2} ( \bar{L}^{(n)} \ast \bar{L}^{(m)} - \bar{L}^{(m)} \ast \bar{L}^{(n)} ) \; .
\eez
Differentiation of (\ref{LX}) yields
\bez
  L_{\theta_{m,n}} = X_{\theta_{m,n}} \ast \pa \ast X^{-1} + X \ast \pa \ast (X^{-1})_{\theta_{m,n}}
 + {1 \over 2} (X_{t_m} \ast \pa \ast(X^{-1})_{t_n} - X_{t_n} \ast \pa \ast (X^{-1})_{t_m})
\eez
and, by use of our previous result, (\ref{LX}), (\ref{XncKPh}), and (\ref{ncKPh}) (which was
already shown to be equivalent to (\ref{XncKPh})),
\bez
  L_{\theta_{m,n}} = [ X_{\theta_{m,n}} \ast X^{-1} , L ]
   + {1 \over 2} ( L_{t_m} \ast \bar{L}^{(n)} - L_{t_n} \ast \bar{L}^{(m)} ) \; .
\eez

Let us assume that (\ref{dthetaX}) holds.
With the help of (\ref{XncKPh}), this equation can be written in the form (\ref{dthetaXL}).
Using the last formula, it is then easy to see that (\ref{L_theta}) is satisfied.

Now let us assume that (\ref{L_theta}) holds. Combining it with the above expression for $L_{\theta_{m,n}}$
leads to
\bez
  [ X_{\theta_{m,n}} \ast X^{-1} - W^{(m,n)}, L ] + {1 \over 2} ( L_{t_m} \ast L^n - L_{t_n} \ast L^m) = 0 \; .
\eez
With the help of $L_{t_n} = - [\bar{L}^{(n)},L]$, we find
\bez
 [ X_{\theta_{m,n}} \ast X^{-1} - W^{(m,n)} + {1 \over 2} ( \bar{L}^{(n)} \ast L^m - \bar{L}^{(m)} \ast L^n), L ] = 0
\eez
and, with the definition (\ref{Wmn}),
\bez
  [ X_{\theta_{m,n}} \ast X^{-1} + \tilde{W}^{(m,n)}, L ] = 0
\eez
where
\bez
 \tilde{W}^{(m,n)} = {1 \over 2} (\bar{L}^{(n)} \ast L^m - \bar{L}^{(m)} \ast L^n)_{<0}
                   = \bar{W}^{(m,n)} - {1 \over 2} [\bar{L}^{(m)},\bar{L}^{(n)}] \; .
\eez
Multiplying by $X^{-1}$ from the left and by $X$ from the right, and using (\ref{LX}), leads to
\bez
  [X^{-1} \ast (X_{\theta_{m,n}} + \tilde{W}^{(m,n)} \ast X)]_x = 0
\eez
and thus
\bez
   X_{\theta_{m,n}} = - \tilde{W}^{(m,n)} \ast X + X \ast C_{m,n}
\eez
where $(C_{m,n})_x = 0$. Next we insert this expression in the integrability conditions
$X_{\theta_{m,n} t_r} - X_{t_r \theta_{m,n}} = 0$ and use (\ref{XncKPh}).
With the help of (\ref{L_theta}), which in terms of $\tilde{W}$ reads
\bez
   L_{\theta_{m,n}}
 = - [\tilde{W}^{(m,n)},L] + {1 \over 2} (L_{t_m} \ast \bar{L}^{(n)} - L_{t_n} \ast \bar{L}^{(m)}) \, ,
\eez
and the $\theta$-$t$-integrability conditions
\bez
   \tilde{W}^{(m,n)}{}_{t_r} - \bar{L}^{(r)}{}_{\theta_{m,n}} - [\tilde{W}^{(m,n)} , \bar{L}^{(r)}]
 = {1 \over 2} ( \bar{L}^{(r)}{}_{t_n} \ast \bar{L}^{(m)} - \bar{L}^{(r)}{}_{t_m} \ast \bar{L}^{(n)} )
\eez
we finally obtain $(C_{m,n})_{t_r} = 0$, i.e., the coefficients of $C_{m,n}$ are only
allowed to depend on the $\theta$'s. Finally, we make use of the $\theta$-$\theta$-integrability
conditions:
\bez
 \lefteqn{\tilde{W}^{(m,n)}{}_{\theta_{r,s}} - \tilde{W}^{(r,s)}{}_{\theta_{m,n}}
 - [\tilde{W}^{(m,n)} , \tilde{W}^{(r,s)}]
 - {1 \over 2} (\tilde{W}^{(m,n)}{}_{t_r} \ast \bar{L}^{(s)} }\hspace*{2.5cm} && \nonumber \\
 & & - \tilde{W}^{(m,n)}{}_{t_s} \ast \bar{L}^{(r)} - \tilde{W}^{(r,s)}{}_{t_m} \ast \bar{L}^{(n)}
   + \tilde{W}^{(r,s)}{}_{t_n} \ast \bar{L}^{(m)}) = 0 \; .
\eez
With their help, $X_{\theta_{m,n} \theta_{r,s}} - X_{\theta_{r,s} \theta_{m,n}} = 0$ leads to
\bez
  (C_{m,n})_{\theta_{r,s}} - (C_{r,s})_{\theta_{m,n}} - [C_{m,n} , C_{r,s}] = 0
\eez
which implies $C_{m,n} = C^{-1} \, C_{\theta_{m,n}}$ (without $\ast$ since $C$ is independent
of the $t_r$). Now
\bez
   X_{\theta_{m,n}} = - \tilde{W}^{(m,n)} \ast X
\eez
is achieved with the gauge transformation $X \to X C$, which does not affect (\ref{XncKPh}).
\hfill $\blacksquare$
\vskip.2cm

As a consequence of this theorem and Theorem~\ref{theorem:XncKPh}, the xncKP hierarchy
can be defined alternatively by (\ref{XncKPh}) and (\ref{dthetaX}).
Using (\ref{dthetaX}), it is easily verified that the Baker-Akhiezer function
$\psi$, which was shown to satisfy the first two equations of the linear system
(\ref{xncKP_ls}), also satisfies the last equation of (\ref{xncKP_ls}).

\section{Bilinear identities}
\label{sec:bil}
\setcounter{equation}{0}
The classical KP hierarchy can be expressed equivalently in terms of so-called bilinear identities
(see \cite{BBT03,Dick03}, for example).
In this section we prove a corresponding result for the xncKP hierarchy.
Furthermore, we recall the route towards the introduction of the $\tau$-function of the KP hierarchy
and discuss briefly problems one has to face in an attempt to find a generalization.
\vskip.1cm

Let $X$ be the dressing operator introduced in section~\ref{sec:B-A}.
Differentiation of the identity $X^{-1} \ast X = 1$ with respect to $\theta_{m,n}$,
using (\ref{dtheta_star}), (\ref{XncKPh}), and (\ref{dthetaXL}), leads to
\be
   (X^{-1})_{\theta_{m,n}} &=& \frac{1}{2} X^{-1} \ast
   \Big( -( \bar{L}^{(m)} \ast L^n - \bar{L}^{(n)} \ast L^m )_{<0}
   + [ \bar{L}^{(m)} , \bar{L}^{(n)} ]_\ast \Big) \nonumber \\
   &=& \frac{1}{2} X^{-1} \ast ( \bar{L}^{(n)} \ast L^{(m)} - \bar{L}^{(m)} \ast L^{(n)} )_{<0}
\ee
and thus
\be
  \psi^\ast_{\theta_{m,n}}
  &=& ((X^{-1})^\dag)_{\theta_{m,n}} \ast e^{-\xi}
      - \frac{1}{2} ( \bar{L}^{(m)} \, \la^n
      - \bar{L}^{(n)} \, \la^m )^\dag \ast \psi^\ast \nonumber \\
  &=& -((X^{-1})_{\theta_{m,n}})^\dag \ast e^{-\xi}
      + \frac{1}{2} ( L^m \ast \bar{L}^{(n)} - L^n \ast \bar{L}^{(m)} )^\dag
        \ast \psi^\ast \nonumber \\
  &=& -((X^{-1})_{\theta_{m,n}})^\dag \ast e^{-\xi}
      + \frac{1}{2} ( \bar{L}^{(n)} \ast L^{(m)} - \bar{L}^{(m)} \ast L^{(n)}
      - [ L^{(m)} , L^{(n)} ]_\ast )^\dag \ast \psi^\ast \nonumber \\
  &=& W^{(m,n)}{}^\dag \ast \psi^\ast
      - {1 \over 2} [ L^{(m)},L^{(n)} ]_\ast^\dag \ast \psi^\ast
\ee
with the help of (\ref{ncKPls_ast}).
\vskip.1cm

We define the \emph{residue} $\mathrm{res}_\la$ of a formal series in $\lambda$
as the coefficient of the $\la^{-1}$ term.\footnote{One often finds
the operation of taking the residue equivalently defined as $\oint d \lambda/(2 \pi i)$
with a contour around $|\lambda| = \infty$ \cite{BBT03}.}
Some useful relations are derived below. We follow the treatment of the commutative case
in \cite{BBT03} (see also \cite{Dick03}).

\begin{lemma} Let $A = \sum_j a_j \, \pa^j$, $B = \sum_k b_k \, \pa^k$ be (formal)
pseudo-differential operators. Then
\be
   \mathrm{res}_\la \Big( ( A \ast e^{\la x}) \ast (B \ast e^{-\la x})^\dag \Big)
 = \mathrm{res}( A \ast B^\dag ) \; .
\ee
\end{lemma}
{\em Proof:} Evaluation of the left hand side yields
\bez
 \mathrm{res}_\la \Big( ( A \ast e^{\la x}) \ast (B \ast e^{-\la x})^\dag \Big)
 &=& \mathrm{res}_\la \Big( \sum_{j,k} (-1)^k \la^{j+k} \, a_j \ast e^{\la x}
     \ast e^{-\la x} \ast b_k^\dag \Big ) \\
 &=& \sum_k (-1)^k \, a_{-k-1} \ast b^\dag_k \, .
\eez
This equals the right hand side since
\bez
   \mathrm{res}(A \ast B^\dag)
 = \mathrm{res} \Big( \sum_j a_j \ast \pa^j \sum_k (-\pa)^k \ast b_k^\dag \Big)
 = \sum_k(-1)^k a_{-k-1} \ast b_k^\dag
\eez
using $\mathrm{res}([\pa^j,f]) = 0$ for all $j \in \mathbb{Z}$ and functions $f(x)$.
\hfill $\blacksquare$

\begin{theorem} {\bf (bilinear identities).}
For all $i_1,\ldots,i_m\in \mathbb{Z}$, with $i_k \geq 0$ and
$j_{1,2}, \ldots, j_{m,n} \in \mathbb{Z}$ with $m<n$ and $j_{k,l} \geq 0$ :
\be
 \mathrm{res}_\la \Big( ( \pa^{i_1}_{t_1} \cdots \pa^{i_m}_{t_m} \pa^{j_{1,2}}_{\theta_{1,2}}
 \cdots \pa^{j_{m,n}}_{\theta_{m,n}} \psi ) \ast \psi^\ddag \Big) = 0    \label{blid}
\ee
with $\psi^\ddag = (\psi^\ast)^\dag$
holds as a consequence of the xncKP hierarchy equations.
\end{theorem}
{\em Proof:} Since $\psi_{t_k} = L^{(k)} \ast \psi$, $\psi_{\theta_{m,n}} = W^{(m,n)} \ast \psi$,
and $L^{(k)}$, $W^{(m,n)}$ are polynomials in $\pa$, it is sufficient to prove the case $i_1 = i \geq 0$,
$i_2 = \cdots = j_{m,n} = 0$. With the help of the preceding Lemma we find
\bez
 \mathrm{res}_\la \Big( (\pa^i \psi) \ast \psi^\ddag \Big)
 &=& \mathrm{res}_\la \Big( (\pa^i X \ast e^\xi) \ast ((X^\dag)^{-1} \ast e^{-\xi})^\dag \Big) \\
 &=& \mathrm{res}_\la \Big( (\pa^i X \ast e^{\la x}) \ast ((X^\dag)^{-1} \ast e^{-\la x})^\dag \Big) \\
 &=& \mathrm{res}(\pa^i X \ast X^{-1}) = \mathrm{res}(\pa^i) = 0 \; .
\eez
\hfill $\blacksquare$
\vskip.2cm

Next we prove the converse: the bilinear equations (\ref{blid}) imply the xncKP hierarchy
equations.

\begin{theorem}
Let
\bez
 \psi := X \ast e^\xi  \qquad
 \psi^\ddag := ( X^\star \ast e^{-\xi} )^\dag
\eez
satisfy the bilinear equations (\ref{blid}), where
\bez
 X := 1 + \sum_{i=1}^\infty w_i(t,\theta) \, \pa^{-i}  \qquad
 X^\star := 1 + \sum_{i=1}^\infty w^\star_i(t,\theta) \, (-\pa)^{-i}
\eez
are formal series with functions $w_i$ and $w^\star_i$.
Then $X^\star = (X^\dag)^{-1}$ and
\bez
   X_{t_n} = - \bar{L}^{(n)} \ast X  \, , \quad
   X_{\theta_{mn}} = \frac{1}{2} ( \bar{L}^{(m)} \ast L^n - \bar{L}^{(n)} \ast L^m )_{<0} \ast X
\eez
where $L = X \ast \pa \ast X^{-1}$. Hence $\psi$ is the Baker-Akhiezer function of
the (extended) ncKP hierarchy.
\end{theorem}
{\em Proof:} From the above definitions we obtain
\bez
 \psi = \Big( 1 + \sum_{i=1}^\infty w_i \, \la^{-i} \Big) \ast e^\xi  \qquad
 \psi^\ddag = e^{-\xi} \ast \Big( 1 + \sum_{i=1}^\infty (w^\star)^\dag_i \, \la^{-i} \Big) \; .
\eez
Let us assume that $\mathrm{res}_\la \Big( (\pa^i \psi) \ast \psi^\ddag \Big) = 0$ for all $i \geq 0$.
With the help of the above Lemma this yields
\bez
 \mathrm{res}(\pa^i X \ast (X^\star)^\dag) = \mathrm{res}_\la \Big( (\pa^i X\ast e^\xi)\ast
(X^\star\ast e^{-\xi})^\dag \Big) = \mathrm{res}_\la \Big( (\pa^i\psi)\ast \psi^\ddag \Big) = 0 \; .
\eez
Since by construction $X \ast (X^\star)^\dag = 1+Y$ with $Y=Y_{<0}$, the last equation implies
$\mathrm{res}(\pa^i Y) = 0$ for all $i \geq 0$ and thus $Y=0$.
It follows that $X^\star = (X^\dag)^{-1}$.

The proof that the ncKP hierarchy equations $X_{t_n} = -\bar{L}^{(n)} \ast X$ hold,
can be carried out in the same way as in the commutative case \cite{BBT03,Dick03}.
With their help the additional equations of the extended hierarchy can be derived as
follows. We find
\bez
     X_{\theta_{m,n}} \ast e^\xi
 &=& (X \ast e^\xi)_{\theta_{m,n}}
     - {1 \over 2} ( X_{t_m} \, \pa^n - X_{t_n} \, \pa^m ) \ast e^\xi \\
 &=& (X\ast e^\xi)_{\theta_{m,n}}
     + {1 \over 2} (\bar{L}^{(m)} \ast L^n - \bar{L}^{(n)} \ast L^m) \ast X \ast e^\xi
\eez
and thus
\bez
 \lefteqn{ ( X_{\theta_{m,n}} - {1 \over 2} (\bar{L}^{(m)} \ast L^n
   - \bar{L}^{(n)} \ast L^m)_{<0} \ast X) \ast e^\xi } \hspace{2cm} \\
 &=& \Big( \frac{\pa}{\pa \theta_{m,n}} + {1 \over 2} (\bar{L}^{(m)} \ast L^n
     - \bar{L}^{(n)} \ast L^m)_{\geq 0} \Big) \ast X \ast e^\xi  \, .
\eez
By application of (\ref{blid}),
\bez
 0 &=& \mathrm{res}_\la \Big( \Big[ \pa^i(\frac{\pa}{\pa \theta_{m,n}} + {1 \over 2} (\bar{L}^{(m)} \, L^n
    - \bar{L}^{(n)} \, L^m)_{\geq 0}) \ast X \ast e^\xi \Big] \ast (X^\star \ast e^{-\xi})^\dag \Big) \\
   &=&  \mathrm{res}_\la \Big( [\pa^i( X_{\theta_{m,n}} - {1 \over 2} (\bar{L}^{(m)} \, L^n
    - \bar{L}^{(n)} \, L^m)_{<0} \ast X) \ast e^\xi] \ast (X^\star \ast e^{-\xi})^\dag \Big)
\eez
so that, using the Lemma,
\bez
  \mathrm{res} \Big( \pa^i( X_{\theta_{m,n}} - {1 \over 2} (\bar{L}^{(m)} \, L^n
   - \bar{L}^{(n)} \, L^m)_{<0} \ast X) \ast X^{-1} \Big) = 0
\eez
for all $i \geq 0$. This implies (\ref{dthetaXL}).
\hfill $\blacksquare$
\vskip.2cm

We have shown that the bilinear identities (\ref{blid}) are equivalent to the xncKP hierarchy
equations.\footnote{We refer to appendix C for a concrete evaluation of bilinear identities.}
They are generated by formal Taylor expansion of
\be
 \mathrm{res}_\la \Big( \psi(t + a,\theta + \alpha,\la) \ast \psi^\ddag(t,\theta,\la) \Big) = 0
\ee
where $t+a$ and $\theta + \alpha$ stand for the collection of $t_n + a_n$, respectively
$\theta_{m,n} + \alpha_{m,n}$, with arbitrary constants $a_n$ and $\alpha_{n,m} = - \alpha_{m,n}$.
With the shift $t \to t-a$, this becomes
\be
 \mathrm{res}_\la \Big( \psi(t,\theta+\alpha,\la) \ast \psi^\ddag(t-a,\theta,\la) \Big) = 0 \; .
   \label{blid_shift}
\ee
A similar shift in $\theta$ affects the $\ast$-product. To make this manifest, let
us write $\ast_\theta$ instead of $\ast$. Using (\ref{BAfunction}), we find
\be
   \psi(t,\theta + \alpha,\la)
 = \hat{w}(t,\theta+\alpha,\la) \ast_{\theta+\alpha} e^\xi
 = \hat{w}(t+\alpha(\la),\theta+\alpha,\la) \ast_\theta e^\xi
\ee
with
\be
    \alpha_m(\la) = \frac{1}{2} \sum_{n=1}^\infty \alpha_{m,n} \, \la^n \; .
\ee
Here we used the definition of the $\ast$-product with $P = P_\theta + P_\alpha$ and
\be
   P_\alpha w(t,\theta,\lambda) \otimes e^\xi
 = 2 \, \Big( \sum_{m=1}^\infty \alpha_m(\lambda) \, w(t,\theta,\lambda)_{t_m} \Big) \otimes e^\xi
\ee
which implies
\be
   e^{P_\alpha/2} w(t,\theta,\lambda) \otimes e^\xi
 = \Big( e^{ \sum_{m=1}^\infty \alpha_m(\lambda) \, \pa_{t_m} } w(t,\theta,\lambda) \Big) \otimes e^\xi
 = w(t+\alpha(\lambda),\theta,\lambda) \otimes e^\xi \; .
\ee
Using (\ref{BAfunction}) and (\ref{BA^ast}), equation (\ref{blid_shift}) can now be turned into
\be
 \mathrm{res}_\la \Big( \hat{w}(t+\alpha(\la),\theta+\alpha,\la) \ast
 \hat{w}^\ast(t-a,\theta,\la) \, e^{-\xi(-a,\la)} \Big) = 0   \label{res_id}
\ee
where $\xi(-a,\la)$ is obtained from $\xi$ by replacing $t_n$ with $-a_n$ for all $n \geq 1$.
\vskip.1cm

Choosing $a = [\la_1^{-1}] = ({1 \over \la_1},{1 \over 2\la_1^2},{1 \over 3\la_1^3},\ldots)$,
we find
\be
    e^{-\xi(-[\la_1^{-1}],\la)} = {1 \over 1 - \la/\la_1}
\ee
and, with the help of the residue identity (see \cite{Dick03}, for example),
\be
   \mathrm{res}_\lambda \Big( {f(\lambda) \over 1 - \lambda/\lambda'} \Big)
 = \lambda' \, f(\lambda')_{<0}
\ee
for a formal series $f(\la) = \sum_{i=-\infty}^\infty f_i \, \la^{-i}$. Here $f(\lambda')_{<0}$
denotes the part of $f(\lambda)$ which only contains negative powers of $\lambda$, with $\lambda$
replaced by $\lambda'$. With our special choice of $a$, (\ref{res_id}) becomes
\be
 \Big( \hat{w}(t+\alpha(\la_1),\theta+\alpha,\la_1) \ast \hat{w}^\ast(t-[\la_1^{-1}],\theta,\la_1) \Big)_{<0}
  = 0 \; .    \label{res_id1}
\ee
Alternatively, choosing $a = [\la_1^{-1}] + [\la_2^{-1}]$, we find
\be
   \mathrm{res}_\lambda \Big( f(\lambda) \, e^{-\xi(-a,\lambda)} \Big)
 = {\lambda_1 \lambda_2 \over \lambda_2 - \lambda_1} [ f(\lambda_1)_{<0} - f(\lambda_2)_{<0} ]
\ee
by partial fraction decomposition. Now (\ref{res_id}) yields
\be
  && \Big( \hat{w}(t+\alpha(\la),\theta+\alpha,\la)
   \ast \hat{w}^\ast(t-[\la_1^{-1}]-[\la_2^{-1}],\theta,\la) \Big)_{<0} \Big|_{\lambda = \lambda_1}
    \nonumber \\
 &=& \Big( \hat{w}(t+\alpha(\la),\theta+\alpha,\la)
   \ast \hat{w}^\ast(t-[\la_1^{-1}]-[\la_2^{-1}],\theta,\la) \Big)_{<0} \Big|_{\lambda = \lambda_2}
   \; .  \label{res_id2}
\ee
In the non-deformed commutative case, we have
$(\hat{w}(t,\la) \, \hat{w}^\ast(t-a,\la))_{<0} = \hat{w}(t,\la) \, \hat{w}^\ast(t-a,\la) - 1$,
so that equations (\ref{res_id1}) and (\ref{res_id2}) combine to
\be
    {\hat{w}(t, \la_1) \over \hat{w}(t - [\la_2^{-1}],\la_1) }
  = {\hat{w}(t, \la_2) \over \hat{w}(t - [\la_1^{-1}], \la_2) }
\ee
which is solved by $\hat{w}(t,\la) = \tau(t-[\la^{-1}])/\tau(t)$
with a function $\tau$ (see also \cite{BBT03,Dick03}, for example).
In the noncommutative case, however, this does not work. We have seen, however,
that many related results indeed generalize to the noncommutative
setting.

\section{Explicit formulae for the xncKP hierarchy equations}
\label{sec:recursion}
\setcounter{equation}{0}
In this section we derive explicit expressions for the equations of the xncKP hierarchy
in terms of $\phi$ and its derivatives. A crucial role in the derivation is played by
the deformation equations (\ref{L_theta}), which imply
\be
     \phi_{\theta_{m,n}}
 = \frac{1}{2} \, \mathrm{res}( \bar{L}^{(n)} \ast L^{(m)} - \bar{L}^{(m)} \ast L^{(n)} )
       \label{phi_thetamn}
\ee
(see also \cite{DMH04hier}). This can be further elaborated as follows,
\be
     2 \, \phi_{\theta_{m,n}}
 &=& \mathrm{res}( \bar{L}^{(n)} \ast (L^m - \bar{L}^{(m)}) - (L^m - L^{(m)}) \ast L^{(n)} ) \nonumber \\
 &=& \mathrm{res}( \bar{L}^{(n)} \ast L^m - L^m \ast L^{(n)})  \nonumber \\
 &=& \mathrm{res}( \bar{L}^{(n)} \ast L^m - L^{m+n} + L^m \ast \bar{L}^{(n)}) \nonumber \\
 &=& \mathrm{res}( - L^{m+n}  - [ \bar{L}^{(n)} , L^m ]_\ast + 2 \, \bar{L}^{(n)} \ast L^m )
      \nonumber \\
 &=& - \phi_{t_{m+n}} + \phi_{t_m \,t_n} + 2 \, \mathrm{res}(\bar{L}^{(n)} \ast L^m) \; .
     \label{phi_thetamn-res}
\ee
Writing
\be
   \bar{L}^{(n)} = - \sum_{m=1}^\infty \sigma_m^{(n)} \ast L^{-m}  \label{sigma}
\ee
with (matrices of) functions $\sigma_m^{(n)}$ (see also
\cite{Wils81cons,MSS90,Arat95,Kupe00,Vlad04}), we obtain\footnote{Whereas the left hand
side is antisymmetric in $m,n$, this is not manifest for the right hand side.
The symmetric part vanishes as a consequence of the recursion relations which
are derived in the following.}
\be
     \phi_{\theta_{m,n}}
 &=& - \frac{1}{2} ( \phi_{t_{m+n}} - \phi_{t_m \,t_n} )
   - \sum_{i=1}^\infty \sigma_i^{(n)} \ast \mathrm{res}( L^{m-i} ) \nonumber \\
 &=& - \frac{1}{2} ( \phi_{t_{m+n}} - \phi_{t_m \,t_n} ) - \sigma_{m+1}^{(n)}
     - \sum_{i=1}^{m-1} \sigma_{m-i}^{(n)} \ast \phi_{t_i}  \label{phi_thetamn_sigma}
\ee
since $\mathrm{res}(L^{-1}) = 1$ and $\mathrm{res}(L^{-m}) = 0$ for $m > 1$ (see also
appendix D). In particular, we have
\be
    \sigma_1^{(n)} = - \mbox{res}(L^n) = - \phi_{t_n}  \; .  \label{sigma_1^n}
\ee

\begin{lemma}
\label{lemma:sigma_recur1}
The $\sigma_m^{(n)}$, $n>1$, are determined via
\be
   \sigma_m^{(n+1)} = \sigma_{m,x}^{(n)} + \sigma_{m+1}^{(n)} + \sigma_{n+m}^{(1)}
                - \sum_{j=1}^{n-1} \sigma_j^{(1)} \ast \sigma_m^{(n-j)}
                + \sum_{j=1}^{m-1} \sigma_{m-j}^{(n)} \ast \sigma_j^{(1)}
                \label{sigma_recur}
\ee
iteratively in terms of the $\sigma_k^{(1)}$.\footnote{Lemma~\ref{lemma:sigma^1_Schur}
determines the coefficients $\sigma_m^{(1)}$.
See also appendix D for an alternative way of computing these functions.}
\end{lemma}
{\em Proof:} Using (\ref{sigma}) on the right hand side of $L^{n+1} = L \ast L^n$,
and taking the non-negative part, we find
\bez
    L^{(n+1)} = \pa L^{(n)} - \sigma_1^{(n)} - \sum_{m=1}^n \sigma_m^{(1)} \ast L^{(n-m)}
\eez
(see also \cite{Vlad04}, for example). Applying both sides of this equation to $\psi$,
using (\ref{sigma}) and the first two equations of the linear system (\ref{xncKP_ls}),
we obtain the above recursion formula.
\hfill $\blacksquare$

\begin{lemma}
\label{lemma:sigma_recur2}
The $\sigma_m^{(n)}$, $n>1$, are iteratively determined by the formulae
\be
  \sigma^{(n+1)}_m &=& \sigma^{(1)}_{m,t_n} + \sigma^{(n)}_{m+1} + \sigma^{(1)}_{n+m}
  - \sum_{j=1}^{n-1} \sigma^{(1)}_j \ast \sigma^{(n-j)}_m
  + \sum_{j=1}^{m-1} \sigma^{(1)}_j \ast \sigma^{(n)}_{m-j}  \label{sigma_recur2} \\
  \sigma^{(1)}_m &=& - {1 \over m} ( \phi_{t_m} + \sum_{j=1}^{m-1} \sigma^{(1)}_{m-j,t_j}) \; .
   \label{sigma^1_m}
\ee
\end{lemma}
{\em Proof:} Applying (\ref{sigma}) to $\psi$ and using the linear system (\ref{xncKP_ls}),
we obtain
\be
    \psi_{t_n} = ( \lambda^n + \sum_{m=1}^\infty \sigma_m^{(n)} \, \lambda^{-m} ) \, \psi \; .
      \label{psi_tn-sigma}
\ee
The integrability conditions $\psi_{t_m t_n} = \psi_{t_n t_m}$ lead to
$\sigma^{(m)}_{1,t_n} = \sigma^{(n)}_{1,t_m}$ and
\bez
  \sigma^{(m)}_{i,t_n} - \sigma^{(n)}_{i,t_m}
  + \sum_{j=1}^{i-1} [ \sigma^{(m)}_{i-j} , \sigma^{(n)}_j ]_\ast = 0
  \qquad i = 2,3, \ldots \; .
\eez
For $n=1$, this reads\footnote{See also Theorem 8.1.9 in \cite{Kupe00}.
As a consequence, in the commutative (undeformed) case, the functions $\sigma_m^{(1)}$
are conserved densities of the KP hierarchy \cite{Wils81cons,MSS90}.
Kupershmidt (\cite{Kupe00}, p. 128) suggests a notion of `nonabelian conserved density'
as an expression, the $t$-derivative of which can be written as a sum of a total
$x$-derivative and commutators. In this sense, the $\sigma^{(1)}_m$ are common
conserved densities of the ncKP hierarchy.}
\bez
  \sigma^{(m)}_{i,x}  = \sigma^{(1)}_{i,t_m}
  + \sum_{j=1}^{i-1} [ \sigma^{(1)}_j , \sigma^{(m)}_{i-j}]_\ast \; .
\eez
Combining this equation with (\ref{sigma_recur}), leads to (\ref{sigma_recur2}).
Setting $i=n-m$ and summing over $m$ (from 1 to $n-1$) yields
\bez
   \sigma^{(n)}_1 - n \, \sigma^{(1)}_n = \sum_{m=1}^{n-1} \sigma^{(1)}_{n-m,t_m} \; .
\eez
With the help of (\ref{ncKPeqs}) and (\ref{sigma_1^n}), this becomes (\ref{sigma^1_m}).
\hfill $\blacksquare$

\begin{lemma}
\label{lemma:sigma^1_Schur}
\be
  \sigma^{(1)}_n = p_n(-\tilde{\pa}) \phi  \label{sigma^1_n-Schur}
\ee
with $\tilde{\pa} = ( \pa_{t_1}, {1\over 2} \pa_{t_2}, {1 \over 3} \pa_{t_3}, \ldots )$
and the Schur polynomials\footnote{The sum is over all partitions
$(1^{m_1} \, 2^{m_2} \, \ldots \, n^{m_n})$ of $n$, such that
$n = m_1 \, 1 + m_2 \, 2 + \ldots + m_n \, n$ with $m_i \in \mathbb{N} \cup \{ 0 \}$.
See \cite{OSTT88}, for example.}
\be
  p_n(t) = \sum_{m_1 + 2 \, m_2 + \ldots + n \, m_n = n \atop m_i \geq 0}
            \frac{1}{m_1! \cdots m_n!} \, t_1^{m_1} \cdots t_n^{m_n}
                    \label{Schur_combinatorial}
\ee
where $t = (t_1,t_2, \ldots)$.
\end{lemma}
{\em Proof:} The linear recursion formula (\ref{sigma^1_m}) does not feel the
non-commutativity. Hence, as in the commutative case, it leads to (\ref{sigma^1_n-Schur}).
See \cite{Arat95,Vlad04}, for example.
\hfill $\blacksquare$
\vskip.2cm

\noindent
{\em Remark.} The Schur polynomials satisfy
\be
    e^{\xi(t,\lambda)} = \sum_{n=0}^\infty \lambda^n \, p_n(t) \; .  \label{e^xi_Schur}
\ee
As a consequence,
\be
  \sum_{n=1}^\infty \sigma^{(1)}_n \la^{-n} = \phi(t-[\lambda^{-1}]) - \phi(t)
\ee
with $[\la_1^{-1}] = ({1 \over \la_1},{1 \over 2 \la_1^2}, {1 \over 3 \la_1^3},
\ldots)$. Differentiating (\ref{BAfunction}) with respect to $x$ and comparing
the result with (\ref{psi_tn-sigma}), we find
$\hat{w}_x = (\phi(t-[\la^{-1}]) - \phi(t) ) \ast \hat{w}$.
In the commutative case, setting $\phi = (\ln \tau)_x$ then leads to
$\hat{w} = \tau(t-[\la^{-1}])/\tau(t)$.
\hfill $\blacksquare$
\vskip.2cm

We can also evaluate (\ref{phi_thetamn}) in the following alternative way:
\be
   2 \, \phi_{\theta_{m,n}}
 &=& \mathrm{res}( L^m \ast \bar{L}^{(n)} - L^n \ast \bar{L}^{(m)} ) \nonumber \\
 &=&  \mathrm{res}(L^m \ast \bar{L}^{(n)} - L^{m+n} + L^n \ast L^{(m)})  \nonumber \\
 &=&  \mathrm{res}(L^m \ast \bar{L}^{(n)} - L^{m+n} + L^{(m)} \ast L^n - (L^n)_{t_m}) \nonumber \\
 &=&  -\phi_{t_{m+n}} - \phi_{t_n \,t_m} + 2 \, \mathrm{res}(L^m \ast\bar{L}^{(n)}) \; .
      \label{phi_thetamn_alt}
\ee
Instead of (\ref{sigma}), this suggests to write
\be
   \bar{L}^{(n)} = \sum_{m=1}^\infty L^{-m} \ast \eta^{(n)}_m  \label{eta}
\ee
with (matrices of) functions $\eta^{(n)}_m$. In particular,
$\eta^{(n)}_1 = \mathrm{res}(L^n) = \phi_{t_n}$.

\begin{lemma}
\label{lemma:eta_recur}
The following recursion relations hold,
\be
  \eta^{(n+1)}_m &=& - \eta^{(1)}_{m,t_n} + \eta^{(n)}_{m+1}
  + \eta^{(1)}_{n+m} - \sum_{j=1}^{m-1} \eta^{(n)}_{m-j} \ast \eta^{(1)}_j
  + \sum_{j=1}^{n-1} \eta^{(n-j)}_m \ast \eta^{(1)}_j  \label{eta_recursion}  \\
  \eta^{(1)}_m &=& \frac{1}{m} ( \phi_{t_m} + \sum_{j=1}^{m-1} \eta^{(1)}_{m-j, t_j} ) \; .
                    \label{eta^1_i}
\ee
The last equation is solved by
\be
    \eta^{(1)}_n = p_n(\tilde{\pa}) \phi \; .  \label{eta^1_n-Schur}
\ee
\end{lemma}
{\em Proof:} Using the involution defined in section~\ref{sec:B-A}, (\ref{eta}) implies
\be
  L^{(m)\dag} = (L^\dag)^m - \sum_{i=1}^\infty \eta^{(m)\dag}_i \ast (L^\dag)^{-i}
     \label{eta2}
\ee
which applied to $\psi^\ast$, and with the help of (\ref{ncKPls_ast}), leads to
\be
    \psi^\ast_{t_m}
  = ( -\la^m + \sum_{i=1}^\infty \eta^{(m)\dag}_i \la^{-i} ) \ast \psi^\ast \; .
       \label{psiast_tm-eta}
\ee
The integrability conditions of these equation are
\be
  \eta^{(m)}_{i,t_n} - \eta^{(n)}_{i,t_m}
  - \sum_{j=1}^{i-1} [\eta^{(m)}_j , \eta^{(n)}_{i-j} ]_\ast = 0  \label{eta_integrability}
\ee
which for $n=1$ become
\be
  \eta^{(m)}_{i,x} = \eta^{(1)}_{i,t_m}
    + \sum_{j=1}^{i-1} [\eta^{(m)}_{i-j} , \eta^{(1)}_j]_\ast \; . \label{eta^m_i,x}
\ee
Using (\ref{eta2}) on the right hand side of the identity
$(L^\dag)^{m+1} = L^\dag \ast(L^\dag)^m$, leads to
\bez
  (L^{m+1})^\dag = - \pa (L^m)^\dag + \sum_{i=1}^\infty \eta^{(1)\dag}_i \ast (L^{m-i})^\dag \; .
\eez
Taking the non-negative part, results in\footnote{Note that
$(A_{\geq 0})^\dag = (A^\dag)_{\geq 0}$ for every pseudo-differential operator $A$.}
\bez
  L^{(m+1)\dag} = - \pa \, L^{(m)\dag} - \eta^{(m)\dag}_1
  + \sum_{i=1}^m \eta^{(1)\dag}_i \ast L^{(m-i)\dag} \; .
\eez
Acting with the last expression on $\psi^\ast$ and using
(\ref{eta2}) and (\ref{eta^m_i,x}),
we obtain the recursion formula (\ref{eta_recursion}).
Setting $i=n-m$ and summing over $m$ (from 1 to $n-1$), leads to
\bez
   n \, \eta^{(1)}_n = \eta^{(n)}_1 + \sum_{m=1}^{n-1} \eta^{(1)}_{n-m, t_m}
\eez
and thus (\ref{eta^1_i}). (\ref{eta^1_n-Schur}) is an obvious analog of
Lemma~\ref{lemma:sigma^1_Schur}.
\hfill $\blacksquare$
\vskip.2cm

The last Lemma supplies us with explicit expressions for the $\eta^{(n)}_m$.
In particular, we obtain
\be
  \eta^{(1)}_1 &=& \phi_x  \qquad
    \eta^{(1)}_2 = {1 \over 2} \phi_y + {1 \over 2} \phi_{xx} \qquad
    \eta^{(1)}_3 = {1 \over 3} \phi_{t_3} + {1 \over 2} \phi_{xy} + {1 \over 6} \phi_{xxx}
                   \nonumber \\
  \eta^{(1)}_4 &=& {1 \over 4} \phi_{t_4} + {1 \over 3} \phi_{x t_3} + {1 \over 8} \phi_{yy}
    + {1 \over 4} \phi_{xxy} + {1 \over 24} \phi_{xxxx} \nonumber \\
  \eta^{(1)}_5 &=& {1 \over 5} \phi_{t_5} + {1 \over 4} \phi_{x \, t_4} + {1 \over 6} \phi_{y t_3}
  + {1 \over 6} \phi_{xx t_3} + {1 \over 8} \phi_{xyy} + {1 \over 12} \phi_{xxxy}
  + {1 \over 120} \phi_{xxxxx} \nonumber \\
  \eta^{(1)}_6 &=& \frac{1}{6} \phi_{t_6} + \frac{1}{5} \phi_{x t_5} + \frac{1}{8} \phi_{xx t_4}
    + \frac{1}{8} \phi_{y t_4} + \frac{1}{18} \phi_{t_3 t_3} + \frac{1}{6} \phi_{xy t_3}
    + \frac{1}{18} \phi_{xxx t_3} + \frac{1}{48} \phi_{yyy} + \frac{1}{16} \phi_{xx yy}
    \nonumber \\
  & & + \frac{1}{48} \phi_{xxxxy} + \frac{1}{720} \phi_{xxxxxx}  \, .   \label{eta-coeffs}
\ee
\vskip.2cm

\noindent
{\em Remark.} Using (\ref{e^xi_Schur}) and (\ref{eta^1_n-Schur}), we get
\be
   \sum_{i=1}^\infty \eta^{(1)}_i \, \lambda^{-i} = \phi(t+[\la^{-1}]) - \phi(t) \; .
\ee
Differentiation of (\ref{BA^ast}) with respect to $x$ and comparison of the
result with (\ref{psiast_tm-eta}), leads to
$\hat{w}^\ast_x = \hat{w}^\ast \ast ( \phi(t+[\la^{-1}]) - \phi(t) )$.
In the commutative case, setting $\phi = (\ln \tau)_x$ yields the expression
$\hat{w}^\ast = \tau(t+[\la^{-1}])/\tau(t)$.
\hfill $\blacksquare$
\vskip.2cm

There is an involution $\omega$, which relates the $\sigma$- and the $\eta$-coefficients.
It is defined by
\be
  (f_{t_n})^\omega = -(f^\omega)_{t_n} \qquad
  (f_{\theta_{m,n}})^\omega = -(f^\omega)_{\theta_{m,n}} \qquad
  (f \ast g)^\omega = - g^\omega \ast f^\omega \qquad
  \phi^\omega = \phi \; .
\ee
Since this involution maps the $\sigma$-recursion relations (see Lemma~\ref{lemma:sigma_recur2})
into the $\eta$-recursion relations (Lemma~\ref{lemma:eta_recur}), and vice versa,
it follows that
\be
  (\eta^{(n)}_m)^\omega = \sigma^{(n)}_m  \; .
\ee
For example, explicit expressions for the $\sigma^{(1)}_m$ are now obtained in a simple
way by applying the involution $\omega$ to (\ref{eta-coeffs}).
The recursion formula (\ref{eta_recursion}) yields
\be
  \eta^{(n)}_2  = \phi_{t_{n+1}} + \phi_{x \, t_n} - \eta^{(1)}_{n+1}
    - \sum_{i=1}^{n-1} \phi_{t_{n-i}} \ast \eta^{(1)}_i   \label{eta^n_2}
\ee
which, by application of $\omega$, leads to
\be
  \sigma^{(n)}_2 = - \phi_{t_{n+1}} + \phi_{x \, t_n} - \sigma^{(1)}_{n+1}
    - \sum_{i=1}^{n-1} \sigma^{(1)}_i \ast \phi_{t_{n-i}} \; .  \label{sigma^n_2}
\ee
\vskip.1cm

Inserting (\ref{eta}) in (\ref{phi_thetamn_alt}), we obtain
\be
     \phi_{\theta_{m,n}}
 &=& - {1 \over 2} ( \phi_{t_{m+n}} + \phi_{t_n \, t_m} )
    + \sum_{i=1}^\infty \mathrm{res}(L^{m-i}) \ast \eta^{(n)}_i  \nonumber \\
 &=& - {1 \over 2} ( \phi_{t_{m+n}} + \phi_{t_n \, t_m} ) + \eta^{(n)}_{m+1}
        + \sum_{i=1}^{m-1} \phi_{t_i} \ast \eta^{(n)}_{m-i} \; . \label{phi_thetamn_eta}
\ee
Addition of (\ref{phi_thetamn_sigma}) and (\ref{phi_thetamn_eta}) yields
\be
 \phi_{\theta_{m,n}} = - {1 \over 2} \Big( \phi_{t_{m+n}}
  + \sigma^{(n)}_{m+1} - \eta^{(n)}_{m+1}
  + \sum_{i=1}^{m-1} ( \sigma^{(n)}_{m-i} \ast \phi_{t_i} - \phi_{t_i} \ast \eta^{(n)}_{m-i} )
  \Big)    \label{phi_thetamn_tm+n}
\ee
whereas subtraction results in
\be
  \phi_{t_m \, t_n} = \sigma^{(n)}_{m+1} + \eta^{(n)}_{m+1}
  + \sum_{i=1}^{m-1} ( \sigma^{(n)}_{m-i} \ast \phi_{t_i} + \phi_{t_i} \ast \eta^{(n)}_{m-i} )
    \; .    \label{phi_tmtn}
\ee
In particular, from these equations we obtain
\be
 \phi_{\theta_{1,n}} &=& - {1 \over 2} \phi_{t_{n+1}} + {1 \over 2} (\eta^{(n)}_2 - \sigma^{(n)}_2)
       \label{phi_theta_1n} \\
 \phi_{\theta_{2,n}} &=& - {1 \over 2} \phi_{t_{n+2}} + {1 \over 2} (\eta^{(n)}_3 - \sigma^{(n)}_3)
 + {1 \over 2} \{\phi_x,\phi_{t_n}\}_\ast   \label{phi_theta_2n}  \\
 \phi_{\theta_{3,n}} &=& - {1 \over 2} \phi_{t_{n+3}} + {1 \over 2} (\eta^{(n)}_4 - \sigma^{(n)}_4)
 + {1 \over 2} \{\phi_y , \phi_{t_n}\}_\ast
 + {1 \over 2} ( \phi_x \ast \eta^{(n)}_2 - \sigma^{(n)}_2 \ast \phi_x )
\ee
where $\{ \; , \; \}_\ast$ denotes the anti-commutator, and
\be
  (\phi_{t_n})_x &=& \eta^{(n)}_2 + \sigma^{(n)}_2  \label{phi_xt_n} \\
  (\phi_{t_n})_y &=& \eta^{(n)}_3 + \sigma^{(n)}_3 + [\phi_x,\phi_{t_n}]_\ast \; . \label{phi_yt_n}
\ee
Equations (\ref{phi_thetamn_tm+n})-(\ref{phi_yt_n}) are main results of this section.
Note that (\ref{phi_tmtn}) and the last two equations no longer involve $\theta$-derivatives,
although they originated from the deformation equations.
\vskip.2cm

\noindent
{\em Remark.} By inspection of the recursion relations in Lemma~\ref{lemma:sigma_recur2},
one finds that
\be
     \sigma^{(n)}_{m+1} = - \frac{n}{m+n} \, \phi_{t_{m+n}} + \ldots
\ee
where the remaining terms only contain $t_k$-derivatives of $\phi$ with $k<m+n$.
Using the $\omega$-involution to obtain a corresponding expression for $\eta^{(n)}_{m+1}$,
we see that the leading derivative term on the right hand side of (\ref{phi_thetamn_tm+n})
is ${1 \over 2}(n-m) \, \phi_{t_{m+n}}$. Furthermore, the terms with $\phi_{t_{m+n}}$
cancel each other in the combination $\eta^{(n)}_{m+1} + \sigma^{(n)}_{m+1}$. Note, however,
that this expressions contains a term proportional to $\phi_{t_{m+n-1} x}$. This means that
(\ref{phi_xt_n}) is not yet solved for $\phi_{t_n x}$. But for $n>2$ it can indeed always
be solved for $\phi_{t_n x}$ due to the following argument.
Using (\ref{sigma^1_n-Schur}), (\ref{eta^1_n-Schur}), and the combinatorial
formula (\ref{Schur_combinatorial}) for the Schur polynomials, it follows that
\be
  \sigma^{(1)}_{n+1} &=& - {1 \over n+1} \phi_{t_{n+1}} + {1 \over n} \phi_{x \,t_n}
    + {1 \over 2(n-1)} \phi_{y \, t_{n-1}} - {1 \over 2(n-1)} \phi_{xx \, t_{n-1}}
    + \ldots \\
  \eta^{(1)}_{n+1} &=& {1 \over n+1} \phi_{t_{n+1}} + {1 \over n} \phi_{x \, t_n}
    + {1 \over 2(n-1)} \phi_{y \, t_{n-1}} + {1 \over 2(n-1)} \phi_{xx \, t_{n-1}}
    + \ldots
\ee
where the remaining terms only contain derivatives with respect to $t_k$ with $k<n-1$.
With the help of (\ref{eta^n_2}) and (\ref{sigma^n_2}), (\ref{phi_xt_n}) takes the form
$\frac{1}{n}(n-2) \, \phi_{t_n \, x} + \ldots = 0$, where the dots stand for terms
which contain $t_k$-derivatives of $\phi$ with $k<n$ only. For $n>2$, we can thus
solve for $\phi_{t_n}$ with an $x$-integration. In this way one arrives
iteratively at integro-differential expressions for the $\phi_{t_n}$,
which only contain derivatives with respect to $x$ and $y$, and $x$-integrals.
In section~\ref{sec:xncKP} we construct such a representation of the xncKP hierarchy
equations in a different way.
\hfill $\blacksquare$
\vskip.2cm

With the help of the above results, we can further evaluate (\ref{phi_theta_1n}):
\be
 \phi_{\theta_{1,2}} &=& {1 \over 6} (\phi_{t_3} - \phi_{xxx}) - \phi_x \ast \phi_x \; .
  \label{phi_theta12_t3}   \\
 \phi_{\theta_{1,3}} &=& {1 \over 4} \Big( \phi_{t_4} - \phi_{xxy} - 3 \, \{\phi_x,\phi_y\}_\ast
    - [\phi_x , \phi_{xx}]_\ast \Big)
   \label{phi_theta13_t4}    \\
 \phi_{\theta_{1,4}}&=& {3 \over 10} \phi_{t_5} - {1 \over 6} \phi_{xxt_3}- {1 \over 8} \phi_{xyy}
  - {1 \over 120} \phi_{xxxxx} - {2 \over 3} \{\phi_x , \phi_{t_3} \}_\ast
  - {1 \over 12} \{\phi_x , \phi_{xxx}\}_\ast         \nonumber \\
  & & - {1 \over 4} [\phi_x,\phi_{xy}]_\ast + {1 \over 4} [\phi_{xx} , \phi_y]_\ast
      - {1 \over 2} \phi_y{}^{\ast2}  \; . \label{phi_theta14_t5}
\ee
For $n=2$, (\ref{phi_xt_n}) is identically satisfied. For $n=3,4,5$, we obtain
\be
  (\phi_{t_3})_x &=& \frac{1}{4} \Big( 3 \, \phi_{yy} + \phi_{xxxx} - 6 \, [\phi_x,\phi_y]_\ast
  + 6 \, (\phi_x{}^{\ast2})_x \Big)    \label{ncKP13}  \\
  (\phi_{t_4})_x &=& \frac{1}{3} \Big( 2 \, \phi_{y t_3} + \phi_{xxxy}
  - [\phi_x , 4 \, \phi_{t_3} - \phi_{xxx}]_\ast + 3 \, (\{\phi_x , \phi_y\}_\ast)_x \Big)
           \label{ncKP14}  \\
  (\phi_{t_5})_x &=& \frac{1}{216} \Big( 90 \, \phi_{y t_4} + 40 \, \phi_{t_3 t_3}
   + 40 \, \phi_{t_3 xxx} + 45 \, \phi_{xxyy} + \phi_{xxxxxx}
   + 270 \, [\phi_{t_4} , \phi_x]_\ast \nonumber \\
  & & + 60 \, [\phi_{t_3} , \phi_y + 3 \, \phi_{xx} ]_\ast
      + 120 \, \{\phi_{x t_3} , \phi_x \}_\ast
      + 45 \, \{\phi_{yy} , \phi_x \}_\ast  \nonumber \\
  & & + 180 \, \{\phi_{xy} , \phi_y \}_\ast + 60 \, [ \phi_y , \phi_{xxx} ]_\ast
      + 90 \, [ \phi_x , \phi_{xxy} ]_\ast + 15 \, \{ \phi_{xxxx} , \phi_x \}_\ast \Big) \; .
            \label{ncKP15}
\ee
\vskip.1cm

Now let us elaborate (\ref{phi_theta_2n}) and (\ref{phi_yt_n}).
Setting $n=1$ in (\ref{eta_recursion}) yields
\be
 \eta^{(2)}_n = - \eta^{(1)}_{n,x} + 2 \, \eta^{(1)}_{n+1}
    - \sum_{i=1}^{n-1} \eta^{(1)}_i \ast \eta^{(1)}_{n-i}
\ee
and its $\omega$-dual
\be
 \sigma^{(2)}_n = \sigma^{(1)}_{n,x} + 2 \, \sigma^{(1)}_{n+1}
    + \sum_{i=1}^{n-1} \sigma^{(1)}_i \ast \sigma^{(1)}_{n-i} \; .
\ee
Furthermore, setting $m=2$ in (\ref{eta_recursion}), leads to
\be
 \eta^{(n)}_3 = {1 \over 2} (\phi_{y \, t_n} + \phi_{xx \, t_n})
   + \phi_{t_n} \ast \phi_x + \eta^{(n+1)}_2 - \eta^{(1)}_{n+2}
   - \sum_{i=1}^{n-1} \eta^{(i)}_2 \ast \eta^{(1)}_{n-i}
\ee
and a corresponding expression for $\sigma^{(n)}_3$ via $\omega$-involution.
By use of these expressions, (\ref{phi_theta_2n}) with $n=3$ results in
\be
 \phi_{\theta_{2,3}} =
 {1 \over 10} \phi_{t_5} - {1 \over 8} \phi_{xyy} + {1 \over 40} \phi_{xxxxx}
    - {3 \over 4} \phi_y{}^{\ast2} - {1 \over 4} [\phi_x , \phi_{xy}]_\ast
    + {1 \over 4} \{\phi_x , \phi_{xxx}\}_\ast
    + {1 \over 4} \phi_{xx}{}^{\ast2} + \phi_x{}^{\ast3} \, . \quad \label{phi_theta23_t5}
\ee
Equation (\ref{phi_yt_n}) with $n=3$ coincides with (\ref{ncKP14}).

\section{A special form of the xncKP equations}
\label{sec:xncKP}
\setcounter{equation}{0}
With the help of appendices A and B, we can evaluate the right hand side of
(\ref{ncKPeqs}) directly in terms of $\phi$ and its derivatives with respect to $x$
and $y = t_2$.\footnote{Alternatively, we can derive the resulting equations iteratively by
solving equations like (\ref{ncKP13})-(\ref{ncKP15}) for the (highest) $t_n$-derivative
and eliminating $t_k$-derivatives with $2<k<n$ on the right hand sides of the resulting
equations. The presentation in this section allows for a somewhat more direct computation
of the desired form of the equations, in particular with the help of computer algebra.
We actually used the two ways in order to check our results.}
Introducing the abbreviations
\be
       \Phi^{(1)} &:=& \phi_x  \\
       \Phi^{(2)} &:=& \int ( 2 \, [\Phi^{(1)} , \phi_x]_\ast + \Phi^{(1)}_y ) \, dx
                    = \phi_y                                      \label{Phi2}  \\
       \Phi^{(3)} &:=& \int ( 2 \, [\Phi^{(2)} , \phi_x]_\ast + \Phi^{(2)}_y ) \, dx
                    = \int ( 2 \, [\phi_y , \phi_x]_\ast + \phi_{yy} ) \, dx      \label{Phi3}  \\
       \Phi^{(4)} &:=& \int ( 2 \, [ \Phi^{(3)} , \, \phi_x ]_\ast + \Phi^{(3)}_y ) \, dx
                 + 2 \int \{ \phi_x , \Phi^{(2)}_x \}_\ast \, dx
                                          \label{Phi4}  \\
       \Phi^{(5)} &:=& \int ( 2 \, [ \Phi^{(4)} , \phi_x ]_\ast + \Phi^{(4)}_y ) \, dx
             + 2 \int \{ \phi_x , \Phi^{(3)}_x \}_\ast \, dx
             + 2 \int [ \phi_{xx} , \phi_{xy} ]_\ast \, dx   \quad            \label{Phi5}
\ee
we obtain the \emph{(potential) ncKP equation}
\be
    \phi_{t_3}
  = \frac{1}{4} ( \phi_{xxx} + 6 \, \phi_x{}^{\ast 2} + 3 \, \Phi^{(3)} )   \label{ncKPpot}
\ee
and the next two ($n=4,5$) evolution equations of the ncKP hierarchy:
\be
     \phi_{t_4}
 &=& \frac{1}{2} \phi_{xxy} + \{ \phi_x , \phi_y \}_\ast + {1 \over 2} \Phi^{(4)}
     \label{ncKPpot2}  \\
      \phi_{t_5}
 &=& \frac{5}{8} \Big( \frac{1}{10} \phi_{xxxxx} + \phi_{xyy}
      + 2 \, \phi_y{}^{\ast 2} + \{ \phi_x , \phi_{xxx} + \Phi^{(3)} \}_\ast
      + \phi_{xx}{}^{\ast 2} + 4 \, \phi_x{}^{\ast 3}     \nonumber \\
  & & + {1 \over 2} \Phi^{(5)} - ( [ \phi_x , \phi_y ]_\ast )_x \Big) \; . \label{ncKPpot3}
\ee
The quantities $\Phi^{(n)}$, $n>2$, defined above arise by separating integrals from the remaining
terms in the respective ncKP hierarchy equation. They show a certain (imperfect) building law
which is related to the existence of recursion operators for the KP hierarchy, see appendix F.
\vskip.1cm

With the help of (\ref{sigma}), we can also express (\ref{phi_thetamn}) in the form
\be
      \phi_{\theta_{m,n}}
 &=& {1 \over 2} \Big( \sum_{i=1}^{n+1} \sigma_i^{(m)} \ast \mbox{res}(L^{n-i})
    - \sum_{i=1}^{m+1} \sigma_i^{(n)} \ast \mbox{res}(L^{m-i}) \Big) \nonumber \\
 &=& {1 \over 2} \Big( \sigma_{n+1}^{(m)} - \sigma_{m+1}^{(n)}
     - \sum_{i=1}^{n-1} \sigma_i^{(m)} \ast \sigma_1^{(n-i)}
     + \sum_{i=1}^{m-1} \sigma_i^{(n)} \ast \sigma_1^{(m-i)} \Big)
     \label{phi_thetamn_antisym}
\ee
and use Lemma~\ref{lemma:sigma_recur1} and appendix D to obtain expressions
in terms of the $u_k$. Then we use again formulae from appendix A to find
\be
        \phi_{\theta_{1,2}}
  &=& \frac{1}{8} ( \Phi^{(3)} - \phi_{xxx} ) - \frac{3}{4} \phi_x{}^{\ast 2}
                   \label{phi_theta12} \\
        \phi_{\theta_{1,3}}
  &=&   \frac{1}{8} \Phi^{(4)}
      - \frac{1}{8} \phi_{xxy} - \frac{1}{2} \{ \phi_x , \phi_y \}_\ast
      + \frac{1}{4} [ \phi_{xx} , \phi_x ]_\ast
                   \label{phi_theta13}  \\
        \phi_{\theta_{1,4}}
  &=& \frac{1}{32} \Big( 3 \, \Phi^{(5)}
        - 10 \, \{ \phi_x , \phi_{xxx} + \Phi^{(3)} \}_\ast
        - \phi_{xxxxx} - 2 \, \phi_{xyy}
        - 10 \, [ \phi_y , \phi_{xx} ]_\ast
                    \nonumber \\
  & &   + 6 \, [ \phi_{xy} , \phi_x ]_\ast
        - 10 \, \phi_{xx}{}^{\ast 2}
        - 4 \, \phi_y{}^{\ast 2} - 40 \, \phi_x{}^{\ast 3} \Big) \qquad
                    \label{phi_theta14} \\
        \phi_{\theta_{2,3}}
  &=& \frac{1}{32} \Big( \Phi^{(5)}
      + 2 \, \{ \phi_x , \Phi^{(3)} \}_\ast
      + \phi_{xxxxx} - 2 \, \phi_{xyy}
      + 10 \, ( \{ \phi_x , \phi_{xxx} \}_\ast + \phi_{xx}{}^{\ast 2} )    \nonumber \\
  & & + 2 \, [ \phi_y , \phi_{xx} ]_\ast + 10 \, [ \phi_{xy} , \phi_x ]_\ast
      - 20 \, \phi_y{}^{\ast 2} + 40 \, \phi_x{}^{\ast 3} \Big) \; .
                    \label{phi_theta23}
\ee
In accordance with (\ref{phi_thetamn_tm+n}), see also the last remark in
section~\ref{sec:recursion}, we observe the following structure:
\bez
    \mbox{ncKP}^{(n)} = \Phi^{(n)}  \qquad
    \phi_{\theta_{m,n}} = a_{m,n} \, \Phi^{(m+n)} + \ldots
\eez
where the first equation stands for the $n$th ncKP equation (with `time' $t_n$),
$a_{m,n}$ are constants, and the dots represent terms which are local in
$\phi, \Phi^{(k)}$, $k < m+n$. Probably, the recursion operators found in \cite{Dorf+Foka92},
see also appendix F, can be modified or supplemented by further recursion operators to cover
the whole ncKP and also the xncKP hierarchy.\footnote{See also \cite{Oeve+Fuch82,Foka+Sant86,Dorf+Foka92},
for example, for recursion operators of the KP hierarchy.}
They should relate expressions containing the $\Phi^{(n)}$. We already followed another
route towards explicit expressions for the xncKP hierarchy equations in section~\ref{sec:recursion},
which does not attempt to express the xncKP equations solely in terms of $\phi$ and its
$x$-and $y$-derivatives, as well as $x$-integrals.
In not insisting on eliminating derivatives with respect to the other variables $t_n$,
we avoid integrals and often achieve simpler formulae (see (\ref{ncKP15}), however).
For example, the system composed of the potential ncKP equation (\ref{ncKPpot}) and
the deformation equation (\ref{phi_theta12}) can be replaced by the considerably simpler
one consisting of (\ref{phi_theta12_t3}) and
\be
 ( 2 \, \phi_{\theta_{1,2}} + \phi_{t_3})_x - \phi_{yy} + 2 \, [\phi_x,\phi_y]_\ast = 0 \; .
     \label{ncKP_theta12_mix}
\ee

\section{Reductions of the xncKP hierarchy}
\label{sec:red}
\setcounter{equation}{0}
Let $\cal A$ be the algebra of pseudo-differential operators with coefficients which
are differential polynomials in $\{ u_i \}_{i=2}^\infty$. Let ${\cal I}_Q$ be the
(two-sided) differential ideal generated by $Q \in {\cal A}$. The xncKP equations
admit a \emph{reduction} to ${\cal A}_Q = {\cal A}/{\cal I}_Q$ if
\be
    ( {\cal I}_Q )_{t_n} \subset {\cal I}_Q \qquad
    ( {\cal I}_Q )_{\theta_{k,l}} \subset {\cal I}_Q  \qquad
    \forall \, n,k,l \in \mathbb{N}    \label{N-reduction}
\ee
(see also \cite{Dick03}, section~5.2). For an equality up to addition of terms
lying in ${\cal I}_Q$ we write `mod $Q$'.
A special reduction is obtained by setting $Q = \bar{L}^{(N)}$ for some fixed
$N \in \mathbb{N}$, so that $L^N = L^{(N)}$ mod $Q$. This is called \emph{$N$-reduction}
\cite{DKJM83,OSTT88,Stram+Oeve90,MSS90,Kono+Oeve93,Hama03b}
and reduces the KP hierarchy to the Gelfand-Dickey hierarchies (see \cite{Dick03},
for example). Indeed,
\be
     (\bar{L}^{(N)})_{t_n}
   = ( (L^N)_{t_n} )_{<0}
   = ( [ L^{(n)} , L^N ]_\ast )_{<0}
   = ( [ L^{(n)} , \bar{L}^{(N)} ]_\ast )_{<0}   \label{Nred1}
\ee
shows that the first of conditions (\ref{N-reduction}) is satisfied \cite{Dick03}.
Since $L^{(rN)} = (L^{(N)})^r = L^{rN}$ mod $Q$ for all $r \in \mathbb{N}$,
we have $L_{t_{rN}} = 0$ mod $Q$ \cite{Dick03}.
The extended hierarchy obtained by $N$-reduction still contains evolution equations
in those variables $t_n$, for which $N$ does not divide $n$. Writing
\be
    L^{(N)} = \pa^N + v_{N-2} \, \pa^{N-2} + v_{N-3} \, \pa^{N-3} + \ldots + v_0
\ee
with (matrices of) functions $v_i$, the $N$-reduction constraint allows to express the
$u_k$, $k=2,3, \ldots$, as differential polynomials of $v_i$, $i=N-k,\ldots,N-2$
\cite{Dick03}. The ncKP equations are thus reduced to
\be
  (L^{(N)})_{t_n} = [L^{(n)},L^{(N)}]_\ast \quad \mbox{mod} \, Q \qquad \quad
   \forall \, n \in \mathbb{N}, \quad n/N \not\in \mathbb{N} \; .
\ee
The relation\footnote{This expression is obtained by taking the negative part
of (4.24) in \cite{DMH04hier}, and using (\ref{Nred1}).}
\be
   (\bar{L}^{(N)})_{\theta_{m,n}}
 &=& \frac{1}{2} ( L^{(n)} \ast \bar{L}^{(N)} \ast L^{(m)}
      - L^{(m)} \ast \bar{L}^{(N)} \ast L^{(n)}
      + \bar{L}^{(N)} \ast [ L^{(m)} , L^{(n)} ]_\ast )_{<0} \nonumber \\
 & &  + ( [W^{(m,n)} , \bar{L}^{(N)} ]_\ast )_{<0}
\ee
shows that also the second part of (\ref{N-reduction}) holds.
Furthermore, we find $L_{\theta_{kN,lN}} = 0$ mod $Q$ for all $k,l \in \mathbb{N}$,
and, by use of (\ref{ncKPh}),
\be
   L_{\theta_{kN,lN+r}} = \frac{1}{2} \, L_{t_{(k+l)N+r}} \quad \mbox{mod} \, Q
       \label{reduction-theta-t}
\ee
for all $k,l \in \mathbb{N}$ and $r = 1,2,\ldots,N-1$.\footnote{The fact that not all
of the $\theta$-equations are independent from the reduced ncKP equations already
follows from the following argument. Since $\phi$ does not depend on $t_{kN}$,
there is no deformation of the product with respect to the parameters
$\theta_{kN,lN+r}$ (see (\ref{Moyal})). If we drop the dependence on all the
variables $\theta_{kN+r,lN+s}$ and also the associated $\theta$-equations,
the $\ast$-product reduces to the ordinary one. Since
$\phi$ is allowed to depend on the variables $\theta_{kN,lN+r}$, we still have
non-trivial $\theta$-equations in this case, the flows of which commute with those of the
\emph{classical} reduced KP hierarchy. If the latter hierarchy is already complete,
it follows that these $\theta$-equations must be equivalent to (combinations of the)
reduced KP hierarchy equations.}
Hence, after reduction, the fields only depend on the combination
of variables $t_{(k+l)N+r} + \theta_{kN,lN+r}/2$.
\vskip.1cm

As a consequence, $N$-reduction of xncKP hierarchy equations amounts to the
following recipe. $\phi$ is only allowed to depend on $t_{kN+r}$
and $\theta_{kN+r,lN+s}$, where $k,l = 0,1,2, \ldots$ and $r,s = 1,2,\ldots,N-1$.
Derivatives of $\phi$ with respect to $t_{mN}$, $m \in \mathbb{N}$, have to be dropped.
Furthermore, each derivative with respect to a variable $\theta_{kN,lN+r}$ has
to be replaced by $1/2$ times the derivative with respect to $t_{(k+l)N+r}$.
In this way, the equations of the reduced hierarchy, expressed in terms of
$\phi$, are easily obtained from the formulae derived in the previous two sections.
\vskip.1cm

Two examples of $N$-reductions are treated in the following subsections.

\subsection{xncKdV hierarchy}
The $2$-reduction condition
\be
      L^2 = \pa^2 + u     \label{L^2KdV}
\ee
leads to expressions for the variables $u_k$ of the ncKP hierarchy
in terms of the new variable $u$ (see appendix E).
The ncKdV hierarchy is the set of equations
\be
    u_{t_{2n+1}} = (L^2)_{t_{2n+1}} = [L^{(2n+1)},L^2]_\ast \qquad n = 1,2,\ldots \; .
    \label{ncKdVh}
\ee
According to (\ref{reduction-theta-t}),
\be
   \frac{1}{2} u_{t_{2(k+l)+1}} =
   u_{\theta_{2k,2l+1}} = \mbox{res}( \bar{L}^{(2l+1)} \ast L^{(2k)})_x
                        = \mbox{res}(L^{2(k+l)+1})_x \; .
                           \label{u_theta-ncKdV}
\ee
In fact, it is well-known that the KdV hierarchy can be written in this way
\cite{Avra+Schi00,Arba02}. According to Lax (see footnote 3 in \cite{Lax68}),
this form of the KdV hierarchy has first been discovered by Gardner.
Indeed, using (\ref{ncKPh}), which implies $(L^m)_{t_n} = [L^{(n)} , L^m ]_\ast$
for $m,n \in \mathbb{N}$, we find
\be
     \mbox{res}(L^{2l+1})_{t_{2k+1}}
 &=&  \mbox{res}( L^{2l+1} )_{t_{2k+1}}  \nonumber \\
 &=& \mbox{res}( L^{(2k+1)}\ast L^{2l+1}-L^{2l+1}\ast L^{(2k+1)} ) \nonumber \\
 &=& \mbox{res}( ( L^{2k+1}-\bar{L}^{(2k+1)}) \ast L^{2l+1} - \bar{L}^{(2l+1)} \ast
      L^{(2k+1)} ) \nonumber \\
 &=& \mbox{res}( L^{2(k+l+1)} - \bar{L}^{(2k+1)} \ast L^{(2l+1)} - \bar{L}^{(2l+1)} \ast
    L^{(2k+1)}) \nonumber \\
 &=& -\mbox{res}(\bar{L}^{(2k+1)} \ast L^{(2l+1)} + \bar{L}^{(2l+1)} \ast L^{(2k+1)}) \; .
\ee
Since the right hand side is symmetric in $k$ and $l$, this implies
\be
   \mbox{res}(L^{2l+1})_{t_{2k+1}} = \mbox{res}( L^{2k+1} )_{t_{2l+1}}
\ee
which includes (\ref{u_theta-ncKdV}) as a special case (via $l \to 0$ and $k \to k+l$).
The last equation implies the commutativity of the ncKdV flows.
\vskip.1cm

The remaining deformation equations are given by
\be
   u_{\theta_{2k+1,2l+1}}
 = \mbox{res}( \bar{L}^{(2l+1)} \ast L^{(2k+1)} - \bar{L}^{(2k+1)} \ast L^{(2l+1)} )_x
          \label{u_theta_odd}
\ee

Instead of evaluating (\ref{ncKdVh}) and (\ref{u_theta_odd}) (see appendix E), we can
apply more directly the simple reduction recipe, mentioned in the
beginning of this section, to the equations of the xncKP hierarchy expressed
in terms of the potential $\phi$. Note that (\ref{u2pot}) and (\ref{L^2KdV}) imply
\be
           u = 2 \, \phi_x \; .
\ee
Using $\phi_{\theta_{1,2}} = -\phi_{t_3}/2$ and $\phi_{\theta_{1,4}} = - \phi_{t_5}/2$,
the deformation equations (\ref{phi_theta12_t3}) and (\ref{phi_theta14_t5}) reproduce the
first two (potential) ncKdV equations:
\be
   \phi_{t_3} &=& {1 \over 4} \, \phi_{xxx} + {3 \over 2} \, \phi_x \ast \phi_x \\
   \phi_{t_5} &=& {1 \over 16} \, \phi_{xxxxx} + {5 \over 8} \, \{ \phi_x , \phi_{xxx} \}_\ast
  + {5 \over 8} \, \phi_{xx}{}^{\ast 2} + {5 \over 2} \, \phi_x{}^{\ast 3} \; .
\ee
Alternatively, the last equation is also obtained from (\ref{phi_theta23_t5})
by using $\phi_{\theta_{2,3}} = -\phi_{t_5}/2$. Furthermore, from (\ref{phi_theta13_t4})
with $\phi_y = 0 = \phi_{t_4}$ we recover (\ref{u_theta13}), i.e.,
\be
   \phi_{\theta_{1,3}} = - {1 \over 4} [\phi_x , \phi_{xx}]_\ast \; .
\ee

\subsection{xncBoussinesq hierarchy}
The \emph{xncBoussinesq hierarchy} is obtained from the xncKP hierarchy by imposing
the $3$-reduction constraint
\be
      L^3 = \pa^3 + u \, \pa + v       \label{Bouss_con}
\ee
with variables $u$ and $v$. This leads to
\be
 u_2 &=& \frac{1}{3} \, u \qquad
 u_3 = \frac{1}{3} \, ( v - u_x ) \qquad
 u_4 = \frac{1}{9} \, ( 2 \, u_{xx} - u^2 - 3 \, v_x ) \nonumber \\
 u_5  &=&  \frac{1}{27} \, ( 6 \, v_{xx} - 3 \, v \ast u - 3 \, u \ast v - 3 \, u_{xxx}
           + 7 \, u \ast u_x + 5 \, u_x \ast u )  \nonumber \\
 u_6 &=& \frac{1}{81} ( 3 \, u_{xxxx} - 9 \, v_{xxx} - 15 \, u_{xx} \ast u
        - 30 \, u \ast u_{xx} + 21 \, u \ast v_x + 15 \, ( v_x \ast u + u_x \ast v )
        \nonumber \\
     & & + 30 \, v \ast u_x - 45 \, u_x \ast u_x - 9 \, v^{\ast 2} + 5 \, u^{\ast 3} )
           \; .  \label{u-Bouss}
\ee
The equations of the ncBoussinesq hierarchy are given by
\be
    u_{t_n} \, \pa + v_{t_n} = (L^3)_{t_n} = [L^{(n)},L^3]_\ast \qquad n = 2,4,5,7,8, \ldots \; .
\ee
 For $n=2$ this yields
\be
   u_y = 2 \, v_x - u_{xx} \qquad
   v_y = v_{xx} - \frac{2}{3} ( u_{xxx} + u \ast u_x - [ u,v ]_\ast )
\ee
where $y=t_2$. Introducing the potential $\phi$ via (\ref{u2pot}), we have
\be
         u = 3 \, \phi_x
\ee
and the first equation leads to
\be
    v = \frac{3}{2} ( \phi_y + \phi_{xx} )
\ee
which, inserted in the second equation, yields the (potential) \emph{ncBoussinesq equation}
\be
   \phi_{yy} = - \frac{1}{3} \phi_{xxxx} - 2 \, (\phi_x{}^{\ast 2})_x
               - 2 \, [ \phi_y , \phi_x ]_\ast \; .    \label{Bouss}
\ee
 For $n=4$, we find
\be
    u_{t_4} = \frac{2}{3} \, v_{xx} - \phi_{xxxx} + 2 \, \{ \phi_x , v \}_\ast
      - 3 \, ( \phi_x{}^{\ast 2} )_x + [ \phi_x , \phi_{xx} ]_\ast \; .
\ee
With the above expression for $v$, this becomes
\be
    \phi_{t_4} = \frac{1}{3} ( \phi_{xxy} + 3 \, \{ \phi_x , \phi_y \}_\ast
                 + [ \phi_x , \phi_{xx} ]_\ast )  \; .   \label{Bouss2}
\ee
This equation is also obtained from (\ref{phi_theta13_t4}) with the help of
$\phi_{\theta_{1,3}} = - \phi_{t_4}/2$. Moreover, using
$\phi_{\theta_{2,3}} = - \phi_{t_5}/2$ in (\ref{phi_theta23_t5}),
we obtain
\be
  \phi_{t_5}
  = {5 \over 24} \phi_{xyy} - {1 \over 24} \phi_{xxxxx} + {5 \over 4} \phi_y{}^{\ast2}
    + {5 \over 12} [\phi_x , \phi_{xy}]_\ast - {5 \over 12} \{\phi_x , \phi_{xxx}\}_\ast
    - {5 \over 12} \phi_{xx}{}^{\ast2} - {5 \over 3} \phi_x{}^{\ast3} \; .
\ee
Furthermore, setting $\phi_{t_3} = 0$ in (\ref{phi_theta12_t3}) and (\ref{phi_theta14_t5}),
leads to
\be
 \phi_{\theta_{1,2}} &=& - {1 \over 6} \, \phi_{xxx} - \phi_x \ast \phi_x \\
 \phi_{\theta_{1,4}}&=& {3 \over 10} \phi_{t_5} - {1 \over 8} \phi_{xyy}
  - {1 \over 120} \phi_{xxxxx} - {1 \over 12} \{\phi_x , \phi_{xxx}\}_\ast
  - {1 \over 4} [\phi_x,\phi_{xy}]_\ast  \nonumber \\
  & & + {1 \over 4} [\phi_{xx} , \phi_y]_\ast - {1 \over 2} \phi_y{}^{\ast2}  \; .
\ee
Equation (\ref{ncKP13}) in this way reproduces the ncBoussinesq equation (\ref{Bouss}).

\section{$N$-soliton solutions of some xncKP equations}
\label{sec:ncKPsol}
\setcounter{equation}{0}
In this section we present $N$-soliton solutions of some of the xncKP equations, following
\cite{Pani01}. We start with the ncKP equation and the first deformation equation
(\ref{phi_theta12_t3}). It is simpler, however, and equivalent to consider (\ref{ncKP_theta12_mix})
instead of the ncKP equation. Inserting the formal series (see also \cite{Okhu+Wada83})
\be
  \phi = \sum_{n=1}^\infty \epsilon^n \phi^{(n)}   \label{eps-expansion}
\ee
in a parameter $\epsilon$ in both equations, and demanding that the resulting equations are
satisfied order by order in $\epsilon$, leads to
\be
      \phi^{(n)}_{\theta_{1,2}} - {1 \over 6} (\phi^{(n)}_{t_3} - \phi^{(n)}_{xxx})
  &=& - \sum_{r=1}^{n-1} \phi^{(r)}_x \ast \phi^{(n-r)}_x     \label{ncKP_eps1}   \\
     ( 2 \, \phi^{(n)}_{\theta_{1,2}} + \phi^{(n)}_{t_3})_x - \phi^{(n)}_{yy}
  &=& - 2 \sum_{r=1}^{n-1} (\phi^{(r)}_x \ast \phi^{(n-r)}_y - \phi^{(r)}_y \ast \phi^{(n-r)}_x) \; .
      \label{ncKP_eps2}
\ee
For $n=1$, these are linear homogeneous equations which are solved by
\be
  \phi^{(1)} = \sum_{k=1}^N \phi_k   \qquad \quad
  \phi_k = c_k \, e^{\xi(t,p_k)} \ast e^{-\xi(t,q_k)}     \label{phi_k_xi}
\ee
where $N \in \mathbb{N}$, $\xi(t,p_k) = \sum_{r\geq 1} t_r \, p_k^r$, and $c_k,p_k,q_k$ are constants.
A solution of the inhomogeneous $n=2$ equations is given by
\be
  \phi^{(2)} = \sum_{k,l=1}^N \frac{\phi_k \ast \phi_l}{q_k - p_l}
\ee
assuming $p_l \neq q_k$ for all $k,l=1, \ldots N$. A corresponding solution exists
for arbitrary $n \in \mathbb{N}$. Indeed, introducing
\be
 \Phi^{(m,n)} = \frac{\phi_m \ast \phi_{m+1} \ast \cdots \ast \phi_n}{(q_m-p_{m+1})(q_{m+1}-p_{m+2})
 \cdots (q_{n-1}-p_n)}  \qquad   m < n   \label{Phi(m,n)}
\ee
and $\Phi^{(m,m)} = \phi_m$, we find that the equations
\be
    \Phi^{(1,n)}_{\theta_{1,2}} - {1 \over 6} (\Phi^{(1,n)}_{t_3} - \Phi^{(1,n)}_{xxx})
  &=& - \sum_{r=1}^{n-1} \Phi^{(1,r)}_x \ast \Phi^{(r+1,n)}_x     \label{10}  \\
     ( 2 \, \Phi^{(1,n)}_{\theta_{1,2}} + \Phi^{(1,n)}_{t_3})_x - \Phi^{(1,n)}_{yy}
  &=& - 2 \sum_{r=1}^{n-1} (\Phi^{(1,r)}_x \ast \Phi^{(r+1,n)}_y - \Phi^{(1,r)}_y \ast \Phi^{(r+1,n)}_x)
\ee
(cf (\ref{ncKP_eps1}) and (\ref{ncKP_eps2})) are satisfied as a consequence of the
algebraic identities
\be
 \Theta^{(1,n)}_{1,2} - {1 \over 6} (T^{(1,n)}_3 - (T^{(1,n)}_1)^3)
  &=& - \sum_{r=1}^{n-1} (q_r-p_{r+1}) \, T^{(1,r)}_1 \, T^{(r+1,n)}_1  \label{sum-id1} \\
     (2 \, \Theta^{(1,n)}_{1,2} + T^{(1,n)}_3) \, T^{(1,n)}_1 - (T^{(1,n)}_2)^2
  &=& - 2 \sum_{r=1}^{n-1} (q_r-p_{r+1}) (T^{(1,r)}_1 \, T^{(r+1,n)}_2 - T^{(1,r)}_2 \, T^{(r+1,n)}_1)
      \qquad  \label{sum-id2}
\ee
where we used the abbreviations
\be
  T^{(m,n)}_r &=& \sum_{k=m}^n (p_k^r-q_k^r)  \\
  \Theta^{(m,n)}_{r,s} &=& {1 \over 2} \sum_{m\leq k<l \leq n} [(p_k^r - q_k^r)(p_l^s-q_l^s)
   - (p_k^s - q_k^s) (p_l^r - q_l^r)] \nonumber \\
   & & - {1 \over 2} \sum_{k=m}^n (p_k^r q_k^s - p_k^s q_k^r) \; .
\ee
The first part of $\Theta$ is due to the interaction of different solitons (corresponding to
terms $\phi_k \ast \phi_l$), while the second part is `intrinsic', it originates from the $\ast$
in the definition of $\phi_k$ (see (\ref{phi_k_xi})). There is thus a formal analogy with orbital
angular momentum and spin.

It follows that
\be
    \phi^{(n)}
  = \sum_{k_1,\ldots,k_n=1}^N \frac{\phi_{k_1} \ast \phi_{k_2} \ast \cdots \ast \phi_{k_n}}
    {(q_{k_1}-p_{k_2})(q_{k_2}-p_{k_3}) \cdots (q_{k_{n-1}}-p_{k_n})}
\ee
solves (\ref{ncKP_eps1}) and (\ref{ncKP_eps2}).
\vskip.2cm

The fact that we were able to obtain solutions of (\ref{phi_theta12_t3}) and (\ref{ncKP_theta12_mix})
to \emph{all} orders in $\epsilon$ is due to the existence of the identities
(\ref{sum-id1}) and (\ref{sum-id2}) which hold for all $n \in \mathbb{N}$.
Similar identities are associated with all other xncKP equations we have explored so far.
For example, the above $N$-soliton solution of the ncKP equation also solves (\ref{phi_theta13_t4})
as a consequence of the family of identities
\be
  \Theta^{(1,n)}_{1,3} &=& {1 \over 4} \Big( T^{(1,n)}_4 - (T^{(1,n)}_1)^2 \, T^{(1,n)}_2
    - 3 \sum_{r=1}^{n-1} (q_r-p_{r+1}) (T^{(1,r)}_1 \, T^{(r+1,n)}_2
    + T^{(1,r)}_2 \, T^{(r+1,n)}_1) \nonumber \\
  & &  - \sum_{r=1}^{n-1}(q_r - p_{r+1}) (T^{(1,r)}_1 \, (T^{(r+1,n)}_1)^2
    - (T^{(1,r)}_1)^2 \, T^{(r+1,n)}_1) \Big)  \; .
\ee
The existence of such families of algebraic identities is the reason why certain partial
differential equations can be solved via (\ref{eps-expansion}) universally to all orders.
In principle, the argument could be reversed: finding a (suitable) family of identities,
it should be possible to construct an associated partial differential equation which can
be solved with the above method. We intend to address these questions in a separate work.
\vskip.2cm

\noindent
{\em Remark.} The $N$-soliton solution $\phi$ can be written in a more compact form.
Using the bra-ket notation
\be
 \langle p| = ( \tilde{c}_1 e^{\xi(\mathbf{x},p_1)}, \ldots, \tilde{c}_N
 e^{\xi(\mathbf{x},p_N)} )
   \qquad
 |q \rangle = ( \tilde{c}'_1 e^{-\xi(\mathbf{x},q_1)}, \ldots, \tilde{c}'_N
 e^{-\xi(\mathbf{x},q_N})^t
\ee
with $\tilde{c}_k \, \tilde{c}'_k = c_k$ and introducing the $N \times N$ matrix
\be
   B = - \left( \frac{|q \rangle_k \langle p |_l}{q_k - p_l} \right)
\ee
(where $|q \rangle_k$ is the $k$th component of $|q \rangle$) we obtain
\be
 \phi^{(n)} = (-1)^{n-1} \, \langle p | \ast B^{\ast (n-1)} \ast | q \rangle
\ee
and, with the help of the geometric series formula, the following simple
expression for the $N$-soliton solution:
\be
 \phi = \sum_{n=0}^\infty (-1)^n \, \langle p | \ast B^{\ast n} \ast |q \rangle
 = \langle p | \ast(I + B)^{\ast -1} \ast | q \rangle  \; .
\ee
In the commutative case with vanishing deformation, this can be rewritten as
\be
   \phi = \mbox{Tr} ( (I+B)^{-1} B_x ) = ( \mbox{Tr} \ln (I+B) )_x = ( \ln \tau )_x
\ee
leading to Hirota's function $\tau = \det(I+B)$. In the presence of deformation,
there seems to be no analog of the $\tau$-function.
\hfill $\blacksquare$
\vskip.2cm

\noindent
{\em Remark.} Setting $q_k = -p_k$, we obtain $T^{(m,n)}_{2r} = 0$, $\Theta^{(m,n)}_{2r,2s} = 0$,
and $\Theta^{(m,n)}_{2r,2s+1} = {1 \over 2} T^{(m,n)}_{2(r+s)+1}$, in accordance with the
$2$-reduction conditions, see section~\ref{sec:red}. Hence we obtain $N$-soliton solutions of
the xncKdV equations in this way. To find corresponding soliton solutions
of the xncBoussinesq hierarchy requires to set $q_k = \zeta \, p_k$ with $\zeta$ a
primitive third root of unity.
\hfill $\blacksquare$

\section{Conclusions}
\label{sec:concl}
\setcounter{equation}{0}
The importance of the KP hierarchy in physics and many branches of mathematics
is expected to be shared to a considerable extent by its deformations and extensions,
thus in particular by the xncKP hierarchy.
We have therefore started a thorough exploration of the latter.
\vskip.1cm

We generalized a considerable part of the Sato formalism from the
commutative KP case to the noncommutative deformed setting. Indeed, several important
results, like the well-known bilinear residue identities, extend to the xncKP hierarchy.
A noncommutative analog of Hirota's $\tau$-function was not obtained in the present
framework, but corresponding progress has been reported in \cite{SWW04}.
\vskip.1cm

The usual definition of the KP hierarchy and its various generalizations by
formulae such as (\ref{ncKPh}) contains the equations in a rather implicit way.
Quite involved computations are needed to derive explicit expressions for the members
of the hierarchy. The computational expense is even higher in case of the
deformation equations. In section~\ref{sec:recursion} we derived formulae
which greatly facilitate the computation of the xncKP equations.
In particular, via calculations in the xncKP hierarchy framework,
we obtained corresponding formulae for the ncKP equations, which then hold for
an arbitrary noncommutative associative product $\ast$. In this way,
information about the classical (nc)KP hierarchy is obtained via an
intermediate step into its Moyal-deformation and extension.
\vskip.1cm

We also considered $N$-reductions of the xncKP hierarchy. In particular, the
corresponding discussion in section~\ref{sec:red} demonstrated that one can even learn
something about \emph{classical} sub-hierarchies from consideration of the xncKP system.
This should provide some motivation to study more of the various facets of reductions
in this framework. We certainly only touched upon this subject. In particular,
there are generalizations of $N$-reductions (see
\cite{Kono+Stra92,Chen92a,Chen92b,Arat95,Kono+Oeve93,Lori+Willo97}, for example,
and the references cited there), which can be further generalized to our framework.
\vskip.1cm

$N$-soliton solutions of the ncKP equation were shown to be also solutions of the
first two deformation equations and it is likely that this holds in general.
The emergence of families of algebraic identities in this connection seems to be
an interesting route for further investigations on its own.
\vskip.1cm

Some computations leading to results presented in this work have been carried
out or checked with the help of the computer algebra software FORM
\cite{Verm00,Verm02,Heck00}.
%

\renewcommand{\theequation} {\Alph{section}.\arabic{equation}}

\section*{Appendix A: Formulae for the coefficients of $L$}
\setcounter{section}{1}
\setcounter{equation}{0}
\addcontentsline{toc}{section}{\numberline{}Appendix A: Formulae for the coefficients of $L$}
In this appendix, we derive expressions for the coefficients $u_k$ in terms of
$x$- and $y$-derivatives of $\phi$. The appearance of $x$-integrals in these expressions
is the price one has to pay for the elimination of other $t_n$-derivatives.
The second of equations (\ref{ncKPh}) reads
\be
 L_y = [L^{(2)},L] = [\pa^2 + 2 \, u_2, L]
     = L_{xx} + 2 \, L_x \pa - 2 \, u_{2,x} + 2 \, [ u_2 , \bar{L} ]_\ast
\ee
where $y = t_2$. Inserting the series (\ref{L-u}) leads to
\be
    \sum_{n=1}^\infty u_{n+1,y} \, \pa^{-n}
 &=&  \sum_{n=1}^\infty ( u_{n+1,xx} + 2 \, u_{n+2,x} + 2 \, u_2 \ast u_{n+1} ) \, \pa^{-n} \nonumber \\
 & & - 2 \sum_{m=1}^\infty \sum_{k=0}^\infty (-1)^k {m+k-1 \choose k} u_{m+1}
         \ast {\pa^k u_2 \over \pa x^k} \, \pa^{-m-k}
\ee
with the help of the basic identity
\be
     \pa^{-m} f
   = \sum_{r=0}^\infty  {-m \choose k} \, \frac{\pa^k f}{\pa x^k} \,\pa^{-m-k}
   = \sum_{k=0}^\infty (-1)^k {m+k-1 \choose k} \, \frac{\pa^k f}{\pa x^k} \, \pa^{-m-k}
   \qquad m > 0     \label{pa^-m_id}
\ee
for a function $f$. Hence
\be
  u_{n+1,y} = u_{n+1,xx} + 2 \, u_{n+2,x} + 2 \, [u_2 , u_{n+1}]_\ast
             - 2 \sum_{k=1}^{n-1} (-1)^k { n-1 \choose k} u_{n-k+1} \ast {\pa^k u_2 \over \pa x^k}
\ee
which yields $u_{3,x} = ( u_{2,y} - u_{2,xx} )/2$ and
\be
  u_{n+1,x} = {1 \over 2} (u_{n,y} - u_{n,xx}) - [u_2,u_n]_\ast
              + \sum_{k=1}^{n-2} (-1)^k {n-2 \choose k} u_{n-k} \ast {\pa^k u_2 \over \pa x^k}
\ee
for $n>2$. In particular,
\bez
   u_{3,x} &=& \frac{1}{2} ( u_{2,y} - u_{2,xx} )    \\
   u_{4,x} &=& \frac{1}{2} ( u_{3,y} - u_{3,xx} - 2 \, u_2 \ast u_{2,x} - 2 \, [ u_2 , u_3 ]_\ast ) \\
   u_{5,x} &=& \frac{1}{2} ( u_{4,y} - u_{4,xx} - 4 \, u_3 \ast u_{2,x} + 2 \, u_2 \ast u_{2,xx}
               - 2 \, [ u_2, u_4 ]_\ast ) \\
   u_{6,x} &=& \frac{1}{2} ( u_{5,y} - u_{5,xx} - 6 \, u_4 \ast u_{2,x} + 6 \, u_3 \ast u_{2,xx}
               - 2 \, u_2 \ast u_{2,xxx} - 2 \, [ u_2, u_5 ]_\ast ) \\
   u_{7,x} &=& \frac{1}{2} ( u_{6,y} - u_{6,xx} - 8 \, u_5 \ast u_{2,x} + 12 \, u_4 \ast u_{2,xx}
               - 8 \, u_3 \ast u_{2,xxx} + 2 \, u_2 \ast u_{2,xxxx} \nonumber \\
           & & - 2 \, [ u_2, u_6 ]_\ast ) \; .
\eez
Introducing the potential $\phi$ via $u_2 = \phi_x$ leads to
\be
 u_3 &=& \frac{1}{2} ( \phi_y - \phi_{xx} ) \label{u3-phi} \\
 u_4 &=& \frac{1}{4} ( \Phi^{(3)}
             + \phi_{xxx} - 2 \, \phi_{xy} - 2 \, \phi_x{}^{\ast 2} )    \label{u4-phi} \\
 u_5 &=& \frac{1}{8} ( \Phi^{(4)}  - 3 \, \phi_{yy} - \phi_{xxxx}
             + 3 \, \phi_{xxy} + 2 \, \phi_x \ast \phi_y - 10 \, \phi_y \ast \phi_x
             \nonumber \\
         & & + 8 \, \phi_x \ast \phi_{xx} + 4 \, \phi_{xx} \ast \phi_x )  \label{u5-phi} \\
 u_6 &=& \frac{1}{8} \Big( \frac{1}{2} \Phi^{(5)} - 2 \, \Phi^{(3)}_y
             + \phi_x \ast \Phi^{(3)} - 7 \Phi^{(3)} \ast \phi_x
                     \nonumber \\
         & & + 8 \, \phi_y \ast \phi_{xx} - 2 \, \phi_y{}^{\ast 2} - 3 \, \phi_{xxx} \ast \phi_x
             + 4 \, \phi_x{}^{\ast 3}
             + ( 3 \, \phi_{yy} + \frac{1}{2} \phi_{xxxx}           \nonumber \\
         & & - 2 \, \phi_{xxy} + 9 \, \phi_y \ast \phi_x - \phi_x \ast \phi_y
             - 11 \, \phi_x \ast \phi_{xx} )_x \Big)
\ee
with the $\Phi^{(n)}$ defined in (\ref{Phi2})-(\ref{Phi5}).

\section*{Appendix B: Residues of powers of $L$}
\setcounter{section}{2}
\setcounter{equation}{0}
\addcontentsline{toc}{section}{\numberline{}Appendix B: Residues of powers of $L$}
In this appendix the residues of the first six powers of $L$ are listed.
\bez
   \mathrm{res}(L)   &=& u_2 \\
   \mathrm{res}(L^2) &=& 2 \, u_3 + u_{2,x} \\
   \mathrm{res}(L^3) &=& 3 \, u_4 + 3 \, u_{3,x} + u_{2,xx} + 3 \, u_2^{\ast 2}  \\
   \mathrm{res}(L^4) &=& 4 \, u_5 + 6 \, u_{4,x} + 4 \, u_{3,xx} + u_{2,xxx}
                       + 6 \, ( u_3 \ast u_2 + u_2 \ast u_3 )  \nonumber \\
                   & & + 4 \, u_{2,x} \ast u_2 + 2 \, u_2 \ast u_{2,x}  \\
   \mathrm{res}(L^5) &=& 5 \, u_6 + 10 \, u_{5,x} + 10 \, u_{4,xx}
                       + 5 \, u_{3,xxx} + u_{2,xxxx}
                       + 10 \, ( u_4 \ast u_2 + u_2 \ast u_4 )
                         \nonumber \\
                   & & + 10 \, u_3^{\ast 2} + 10 \, u_{2,x} \ast u_3
                       + 10 \, ( u_{3,x} \ast u_2 + u_2 \ast u_{3,x} )
                       + 10 \, u_2^{\ast 3}     \nonumber \\
                   & &  + 5 ( u_{2,xx} \ast u_2 + u_2 \ast u_{2,xx}
                            + u_{2,x} \ast u_{2,x} ) \\
   \mathrm{res}(L^6) &=& 6 \, u_7 + 15 \, u_{6,x} + 20 \, u_{5,xx}
                       + 15 \, u_{4,xxx} + 6 \, u_{3,xxxx} + u_{2,xxxxx}    \nonumber \\
                   & & + 15 \, (u_5 \ast u_2 + u_2 \ast u_5 )
                       + 20 \, u_{4,x} \ast u_2 + 25 \, u_2 \ast u_{4,x}
                       + 15 \, ( u_4 \ast u_3 + u_3 \ast u_4 )              \nonumber \\
                   & & - 5 \, u_4 \ast u_{2,x} + 20 \, u_{2,x} \ast u_4
                       + 15 \, u_{3,xx} \ast u_2 + 20 \, u_2 \ast u_{3,xx}
                       + 20 \, u_{3,x} \ast u_3   \nonumber \\
                   & & + 10 \, u_3 \ast u_{3,x}
                       + 5 \, u_{3,x} \ast u_{2,x}
                       + 25 \, u_{2,x} \ast u_{3,x}
                       + 10 \, u_3 \ast u_{2,xx} + 15 \, u_{2,xx} \ast u_3  \nonumber \\
                   & & + 20 \, ( u_3 \ast u_2^{\ast 2} + u_2^{\ast 2} \ast u_3 )
                       + 6 \, u_{2,xxx} \ast u_2 + 4 \, u_2 \ast u_{2,xxx}
                       + 9 \, u_{2,xx} \ast u_{2,x}           \nonumber \\
                   & & + 11 \, u_{2,x} \ast u_{2,xx}
                       + 15 \, u_{2,x} \ast u_2^{\ast 2}
                       + 5 \, u_2^{\ast 2} \ast u_{2,x}
                       + 20 \, u_2 \ast u_3 \ast u_2     \nonumber \\
                   & & + 10 \, u_2 \ast u_{2,x} \ast u_2  \, .
\eez
Note that $\mbox{res} (L^n) = n \, u_{n+1} + \ldots$
where the remaining terms only involve the $u_k$ with $k \leq n$.
Hence, $u_{n+1} = \phi_{t_n}/n + \ldots$ by use of (\ref{ncKPeqs}).

\section*{Appendix C: Evaluation of bilinear identities}
\setcounter{equation}{0}
\setcounter{section}{3}
\addcontentsline{toc}{section}{\numberline{}Appendix C: Evaluation of bilinear identities}
Writing $X = \sum_{n=0}^\infty w_n \, \pa^{-n}$ with $w_0=1$ and
$X^{-1}=\sum_{n=0}^\infty v_n \, \pa^{-n}$, we find the coefficients\footnote{These
coefficients are only used in this appendix. They are different from the $v_n$
which appear in section~\ref{sec:red}.}
$v_n$ as differential polynomials in the $w_n$, using the basic formula (\ref{pa^-m_id}):
\be
 1 &=& X \ast X^{-1}
   = \sum_{k,l=0}^\infty w_k \, \pa^{-k} \ast v_l \, \pa^{-l}
   = \sum_{k,l,r=0}^\infty {-k \choose r} w_k \ast (\pa_x^r v_l) \, \pa^{-k-l-r} \nonumber \\
  &=& \sum_{n=0}^\infty \left[ \sum_{m=0}^n \sum_{r=0}^m {-m+r \choose r} w_{m-r}
      \ast (\pa_x^r v_{n-m}) \right]\pa^{-n}
\ee
where $\pa_x^r v_l = \pa^r v_l / \pa x^r$. Hence
\be
   \sum_{m=0}^n \sum_{r=0}^m {-m+r \choose r} w_{m-r} \ast (\pa_x^r v_{n-m}) = \delta_{n,0}
   \qquad   n=0,1,\ldots
\ee
which leads to $v_0 = 1$, $v_1 = -w_1$, and
\be
  v_n = - w_n - \sum_{m=1}^{n-1}\sum_{r=0}^{m-1}(-1)^r {m-1\choose r} w_{m-r} \ast (\pa_x^r v_{n-m})
        \qquad n=2,3,\ldots  \; .
\ee
Furthermore, using $L = X \ast \pa \ast X^{-1} = \pa \, X \ast X^{-1} - X_x \ast X^{-1}$,
we find
\be
  & & \sum_{n=1}^\infty u_{n+1} \, \pa^{-n}
   = L - \pa
   = - X_x \ast X^{-1}  \nonumber  \\
  &=& - w_{1,x} \, \pa^{-1} - \sum_{n=2}^\infty \left[ \sum_{m=1}^{n-1} \sum_{r=0}^{m-1}(-1)^r {m-1 \choose r}
    w_{m-r,x} \ast (\pa_x^r v_{n-m}) + w_{n,x} \right] \pa^{-n}
\ee
from which we read off $u_2 = - w_{1,x}$ and
\be
  u_{n+1} = -w_{n,x} - \sum_{m=1}^{n-1} \sum_{r=0}^{m-1} (-1)^r {m-1 \choose r} w_{m-r,x}
   \ast (\pa_x^r v_{n-m})    \qquad n=2,3,\ldots  \; .
\ee
This determines the $w_k$ in terms of the $u_k$. Setting $u_2 = \phi_x$, we find
\be
   w_1 = - \phi  \qquad
   w_{2,x} = -u_3 + \phi_x \ast \phi \qquad
   w_{3,x} = -u_4 + u_3 \ast \phi - \phi_x \ast w_2 - \phi_x \ast \phi_x \; .
     \label{w_123}
\ee
The Baker-Akhiezer function $\psi$ is then determined via (\ref{BAfunction}).
\vskip.1cm

In order to elaborate the bilinear identities (\ref{blid}), we still need
expressions in terms of the $u_k$ for the coefficients $w^{(\ast)}_n$
introduced in (\ref{hatw^ast}).
Using $X^\dag = \sum_{n=0}^\infty (-\pa)^{-n} \, w_n^\dag$ and
$(X^\dag)^{-1} = \sum_{n=0}^\infty (w^{(\ast)}_n)^\dag \,(-\pa)^{-n}$ with $w^{(\ast)}_0 = 1$,
the coefficients $w^{(\ast)}_n$ are determined by
\be
  1 &=& (X^\dag)^{-1} \ast X^\dag  = \sum_{k,l=0}^\infty (w^{(\ast)}_k)^\dag \, (-\pa)^{-k-l} \, w_l^\dag
     = \sum_{k,l,r=0}^\infty (-1)^{k+l}{-k-l \choose r} (w^{(\ast)}_k)^\dag \ast (\pa_x^r w_l^\dag) \, \pa^{-k-l-r}
       \nonumber \\
    &=& \sum_{n=0}^\infty \sum_{m=0}^n \sum_{r=0}^m (-1)^{n-r} {-n+r \choose r}
         (w^{(\ast)}_{m-r})^\dag \ast (\pa_x^r w_{n-m}^\dag) \, \pa^{-n} \; .
\ee
Reading off the coefficients of powers of $\pa$ and applying the involution ${ }^\dag$,
leads to
\be
 \sum_{m=0}^n \sum_{r=0}^m {n-1 \choose r} (\pa_x^r w_{n-m}) \ast w^{(\ast)}_{m-r} = 0
   \qquad n=1,2,\ldots
\ee
which implies $w^{(\ast)}_1 = -w_1$ and
\be
   w^{(\ast)}_n
 = -w_n - \sum_{m=1}^{n-1} \sum_{r=0}^m {n-1 \choose r} (\pa_x^r w_{n-m}) \ast w^{(\ast)}_{m-r}
   \qquad n = 2,3,\ldots  \; .
\ee
In particular,
\be
 w^{(\ast)}_2 &=& - w_2 + w_1{}^{\ast 2} - w_{1,x} \label{wast_2}   \\
 w^{(\ast)}_3 &=& - w_3 + w_1 \ast w_2 + w_2 \ast w_1 - w_1{}^{\ast 3} - 2 \, w_{2,x}
              + w_1 \ast w_{1,x} + 2 \, w_{1,x} \ast w_1 - w_{1,xx} \, . \qquad
\ee

In terms of $\hat{w}$ and $\hat{w}^\ast$, the bilinear identities take the form
\be
 \mathrm{res}_\la \Big( \pa_{t_1}^{i_1} \cdots \pa_{t_n}^{i_n} \pa_{\theta_{1,2}}^{j_{1,2}}
 \cdots \pa_{\theta_{m,n}}^{j_{m,n}}(\hat{w} \ast e^\xi) \ast e^{-\xi} \ast \hat{w}^\ast\Big) = 0 \; .
   \label{blid_w}
\ee
Let us evaluate some of them. The simplest case is
\be
    \mathrm{res}_\la (\hat{w} \ast \hat{w}^\ast)
  = \sum_{k,l=0}^\infty \mathrm{res}_\la(w_k \ast w^{(\ast)}_l \, \la^{-k-l})
  = w_1 + w^{(\ast)}_1
  \equiv 0  \; .
\ee
Furthermore, using
\be
  \psi_{t_i} = (\hat{w}_{t_i} + \hat{w} \, \la^i) \ast e^\xi     \qquad
  \psi_{t_i t_j} = (\hat{w}_{t_i t_j} + \hat{w}_{t_i} \, \la^j + \hat{w}_{t_j} \, \la^i
   + \hat{w} \la^{i+j}) \ast e^\xi
\ee
we obtain
\be
 \mathrm{res}_\la [ (\hat{w}_{t_i} + \hat{w} \, \la^i) \ast \hat{w}^\ast ] &=& 0 \\
 \mathrm{res}_\la [ (\hat{w}_{t_i t_j} + \hat{w}_{t_i} \, \la^j + \hat{w}_{t_j} \, \la^i
   + \hat{w} \la^{i+j}) \ast \hat{w}^\ast ] &=& 0
\ee
and thus
\be
  0 &=& w_{1,t_i} + \sum_{k=0}^{i+1} w_k \ast w^{(\ast)}_{i-k+1}  \\
  0 &=& w_{1,t_i t_j} + \sum_{k=1}^{i+1} w_{k,t_j} \ast w^{(\ast)}_{i-k+1}
       + \sum_{k=1}^{j+1} w_{k,t_i} \ast w^{(\ast)}_{j-k+1} + \sum_{k=0}^{i+j+1} w_k \ast w^{(\ast)}_{i+j-k+1}
       \; .
\ee
Of particular interest for us is the case where (\ref{blid_w}) only contains a single
$\theta$-derivative, since from it we recover the extension of the ncKP hierarchy. Using
\be
  \psi_{\theta_{i,j}} = \left( \hat{w}_{\theta_{i,j}}
   + {1 \over 2} ( \hat{w}_{t_i} \, \la^j - \hat{w}_{t_j} \, \la^i) \right) \ast e^\xi
\ee
we obtain
\be
 \mathrm{res}_\la \left[ \left(\hat{w}_{\theta_{i,j}} + {1 \over 2} ( \hat{w}_{t_i} \, \la^j
 - \hat{w}_{t_j} \, \la^i) \right) \ast \hat{w}^\ast \right] = 0
\ee
and thus
\be
  w_{1,\theta_{i,j}} + {1 \over 2} \sum_{k=1}^{j+1} w_{k,t_i} \ast w^{(\ast)}_{j-k+1}
   - {1 \over 2} \sum_{k=1}^{i+1} w_{k,t_j} \ast w^{(\ast)}_{i-k+1} = 0
\ee
which is another expression for the tower of ncKP deformation equations
(\ref{phi_thetamn}), respectively (\ref{phi_thetamn_antisym}).
For example, using $w^{(\ast)}_0 = 1$, $w^{(\ast)}_1 = -w_1$, and (\ref{w_123}),
(\ref{wast_2}), (\ref{u3-phi}), (\ref{u4-phi}),
one recovers (\ref{phi_theta12}).

\section*{Appendix D: Computation of the $\sigma$-coefficients}
\setcounter{equation}{0}
\setcounter{section}{4}
\addcontentsline{toc}{section}{\numberline{}Appendix D: Computation of the $\sigma$-coefficients}
Expressions for the coefficients $\sigma_m^{(1)}$ in terms of the $u_k$
can be obtained from (\ref{sigma}) with $n=1$ as follows. Inserting $L^{(1)} = \pa$,
this leads to
\be
    \sum_{j=1}^\infty \Big( u_{j+1} \, \pa^{-j} + \sigma_j^{(1)} \ast L^{-j} \Big) = 0
\ee
and thus
\be
   \sigma_m^{(1)} = - u_{m+1} - \sum_{j=1}^{m-2} \sigma_j^{(1)} \ast (L^{-j})_{-m}
\ee
where the coefficients $(L^{-j})_{-m}$ can be expressed in terms of the coefficients $\ell_k$
of\footnote{$L$ has a unique left inverse as a formal series.
$L^{-1}$ is also a right inverse as a consequence of associativity.}
\be
     L^{-1} = \sum_{i=1}^\infty \ell_i \, \pa^{-i}
\ee
and their $x$-derivatives. One finds
\be
  \ell_1 &=& 1 \, , \quad  \ell_2 = 0 \, , \quad
  \ell_3 = - u_2 \, , \quad \ell_4 = - u_3 + u_{2,x} \nonumber \\
  \ell_{i+1} &=& - \sum_{k=0}^{i-2} (-1)^k \frac{\pa^k u_{i-k}}{\pa x^k}
  - \sum_{j=3}^{i-1} \ell_j \ast
  \sum_{k=0}^{i-j-1} (-1)^k {j+k-1 \choose k} \, \frac{\pa^k u_{i-j-k+1}}{\pa x^k}
  \qquad i>3 \quad
\ee
so that
\bez
   \ell_5 &=& - u_4 + u_{3,x} - u_{2,xx} + u_2^{\ast 2}  \\
   \ell_6 &=& - u_5 + u_{4,x} - u_{3,xx} + u_{2,xxx} - 3 \, u_2 \ast u_{2,x}
              - u_{2,x} \ast u_2 + u_2 \ast u_3 + u_3 \ast u_2  \\
   \ell_7 &=& - u_6 + u_{5,x} - u_{4,xx} + u_{3,xxx} - u_{2,xxxx}
              + 6 \, u_2 \ast u_{2,xx} + u_{2,xx} \ast u_2
              - 3 \, u_2 \ast u_{3,x}    \nonumber \\
          & & - u_{3,x} \ast u_2 + 4 \, u_{2,x} \ast u_{2,x} - u_{2,x} \ast u_3
              - 4 \, u_3 \ast u_{2,x} + u_3{}^{\ast 2} - u_2^{\ast 3} \nonumber \\
          & & + u_2 \ast u_4 + u_4 \ast u_2
\eez
and so forth. Furthermore, with the help of the identity (\ref{pa^-m_id}) one obtains
\be
   L^{-j} = \sum_{i_1,\ldots,i_j=1 \atop k_1,\ldots,k_{j-1} =0}^\infty
   (-1)^{ \sum_{r=1}^{j-1} k_r } \, \ell_{i_1} \,
   \left( \prod_{s=1}^{j-1} { \sum_{r=1}^s (i_r + k_r) -1 \choose k_s } \,
   \frac{\partial^s}{\partial x^s} \ell_{i_{s+1}} \right) \,
   \partial^{-\sum_{r=1}^j i_r - \sum_{r=1}^{j-1} k_r}
\ee
for $j > 1$, which yields
\be
     (L^{-j})_{-m}
 &=& \sum_{ { i_1,\ldots,i_j \geq 1 \atop k_1,\ldots,k_{j-1} \geq 0 }
      \atop i_1 + \ldots i_j + k_1 + \ldots + k_{j-1} = m }
      (-1)^{\sum_{r=1}^{j-1} k_r} \, {i_1+k_1-1 \choose k_1} {i_1+i_2+k_1+k_2-1 \choose k_2} \cdots
         \nonumber \\
 & & \cdots {i_1+\ldots i_{j-1} + k_1 + \ldots + k_{j-1}-1 \choose k_{j-1}} \ell_{i_1}
      \frac{\pa^{k_1} \ell_{i_2}}{\pa x^{k_1}} \cdots \frac{\pa^{k_{j-1}} \ell_{i_j}}{\pa x^{k_{j-1}}} \; .
            \label{L^-j_m}
\ee
Inspection of this formula shows that $(L^{-j})_{-m} = 0$ if $m < j$, $(L^{-j})_{-j} = 1$,
$(L^{-j})_{-j-1} = 0$, and $(L^{-j})_{-j-2} = j \, \ell_3 = - j \, u_2$. Hence
$L^{-j} = \pa^{-j} - j \, u_2 \, \pa^{-j-2} + \ldots$.
Clearly, $(L^{-1})_{-m} = \ell_m$.
\vskip.1cm

The formulae presented above are suitable for evaluation with computer algebra.
In particular, one obtains
\bez
      \sigma_2^{(1)} &=&  - u_3 \\
      \sigma_3^{(1)} &=&  - u_4 - u_2^{\ast 2}  \\
      \sigma_4^{(1)} &=& - u_5 - 2 \, u_3 \ast u_2 - u_2 \ast u_3 + u_2 \ast u_{2,x} \\
      \sigma_5^{(1)} &=& - u_6 - 3 \, u_4 \ast u_2 - u_2 \ast u_4 - 2 \, u_3^{\ast 2}
                    + 3 \, u_3 \ast u_{2,x} + u_2 \ast u_{3,x}
                    - u_2 \ast u_{2,xx} - 2 \, u_2^{\ast 3}   \; . \qquad
\eez
 From (\ref{sigma_recur}) we find, for example,
\bez
     \sigma_2^{(2)} &=& - 2 \, u_4 - u_{3,x} - u_2^{\ast 2} \\
     \sigma_3^{(2)} &=& - 2 \, u_5 - u_{4,x} - 3 \, u_3 \ast u_2 - u_2 \ast u_3
                   - u_{2,x} \ast u_2 + u_2 \ast u_{2,x} \\
     \sigma_4^{(2)} &=& - 2 \, u_6 - u_{5,x} - 5 \, u_4 \ast u_2 - u_2 \ast u_4
                   - 2 \, u_{3,x} \ast u_2 + u_2 \ast u_{3,x}
                   - 3 \, u_3^{\ast 2}  \nonumber \\
               & & + 4 \, u_3 \ast u_{2,x} - u_{2,x} \ast u_3
                   + u_{2,x} \ast u_{2,x} - u_2 \ast u_{2,xx} - 2 \, u_2^{\ast 3} \\
     \sigma_2^{(3)} &=& - 3 \, u_5 - 3 \, u_{4,x} - u_{3,xx} - 3 \, ( u_3 \ast u_2 + u_2 \ast u_3 )
                   + u_2 \ast u_{2,x} - u_{2,x} \ast u_2 \\
     \sigma_3^{(3)} &=& - 3 \, u_6 - 3 \, u_{5,x} - u_{4,xx} - 6 \, u_4 \ast u_2 - 3 \, u_2 \ast u_4
                   - 4 \, u_{3,x} \ast u_2 + u_2 \ast u_{3,x} \nonumber \\
               & & - 3 \, u_3^{\ast 2} + 4 \, u_3 \ast u_{2,x} - u_{2,x} \ast u_3
                   - u_{2,xx} \ast u_2 - u_2 \ast u_{2,xx}
                   + u_{2,x} \ast u_{2,x} - 4 \, u_2^{\ast 3} \\
     \sigma_2^{(4)} &=& - 4 \, u_6 - 6 \, u_{5,x} - 4 \, u_{4,xx} - u_{3,xxx}
                   - 6 \, ( u_4 \ast u_2 + u_2 \ast u_4 )
                   - 4 \, u_{3,x} \, u_2 \nonumber \\
               & & - 2 \, u_2 \ast u_{3,x} - 6 \, u_3^{\ast 2} + 4 \, u_3 \ast u_{2,x}
                   - 4 \, u_{2,x} \ast u_3
                   - u_{2,xx} \ast u_2 - u_2 \ast u_{2,xx} \nonumber \\
               & & + u_{2,x} \ast u_{2,x} - 4 \, u_2^{\ast 3}  \; .
\eez
Note also that $\sigma^{(n)}_1 = - \mbox{res}(L^n)$ with corresponding expressions
listed in appendix B.
The results in this appendix in fact hold for an arbitrary associative product for which
$\pa$ is a derivation. In the commutative case, corresponding expressions for the
$\sigma^{(n)}_m$ can be found in the appendix of \cite{MSS90}.

\section*{Appendix E: Some xncKdV formulae}
\setcounter{equation}{0}
\setcounter{section}{5}
\addcontentsline{toc}{section}{\numberline{}Appendix E: Some xncKdV formulae}
In particular, (\ref{L^2KdV}) leads to
\be
 u_2  &=&  2^{-1} \, u \qquad \quad
 u_3   =  2^{-2} \, (-u_x) \qquad \quad
 u_4   =  2^{-3} \, (u_{xx}-u^{\ast2} ) \nonumber \\
 u_5  &=&  2^{-4} \, ( - u_{xxx}+ 4 \, u \ast u_x + 2 \, u_x \ast u ) \nonumber \\
 u_6  &=&  2^{-5} \, ( u_{xxxx} - 11 \, u \ast u_{xx} - 3 \, u_{xx} \ast u
            - 11 \, u_x \ast u_x + 2 \, u^{\ast 3} ) \nonumber \\
 u_7  &=&  2^{-6} \, ( -u_{xxxxx} + 26 \, u \ast u_{xxx} + 4 \, u_{xxx} \ast u
            + 39 \, u_x \ast u_{xx} + 21 \, u_{xx} \ast u_x
            \nonumber \\
      & &  - 15 \, u^{\ast 2} \ast u_x - 10 \, u \ast u_x \ast u - 5 \, u_x \ast u^{\ast 2} )
             \nonumber \\
 u_8  &=& 2^{-7} \, ( u_{xxxxxx} - 5 \, u_{xxxx} \ast u - 57 \, u \ast u_{xxxx}
           - 34 \, u_{xxx} \ast u_x - 114 \, u_x \ast u_{xxx}
           \nonumber \\
      & &  - 91 \, u_{xx}{}^{\ast 2} + 9 \, u_{xx} \ast u^{\ast 2}
           + 69 \, u^{\ast 2} \ast u_{xx}
           + 32 \, u \ast u_{xx} \ast u + 32 \, u_x{}^{\ast 2} \ast u  \nonumber \\
      & &  + 92 \, u \ast u_x{}^{\ast 2} + 46 \, u_x \ast u \ast u_x - 5 \, u^{\ast 4} ) \; .
\ee
By use of these expressions, we find
\be
   L^{(3)} &=& \pa^3 + \frac{3}{2} u \, \pa + \frac{3}{4} u_x  \\
   L^{(5)} &=& \pa^5 + \frac{5}{2} u \, \pa^3 + \frac{15}{4} u_x \, \pa^2
             + \frac{5}{8} ( 5 \, u_{xx} + 3 \, u^{\ast 2} ) \, \pa
             + \frac{5}{16} ( 3 \, u_{xxx} + 2 \, u_x \ast u + 4 \, u \ast u_x )
               \quad \\
   L^{(7)} &=& \pa^7 + \frac{7}{2} u \, \pa^5 + \frac{35}{4} u_x \, \pa^4
             + \frac{35}{8} ( 3 \, u_{xx} + u^{\ast 2} ) \, \pa^3 \nonumber \\
           & & + \frac{5}{16} ( 35 \, u_{xxx} + 14 \, u_x \ast u + 28 \, u \ast u_x ) \, \pa^2
               \nonumber \\
           & & + \frac{1}{32} ( 161 \, u_{xxxx} + 105 \, u_{xx} \ast u + 245 \, u_x^{\ast 2}
               + 245 \, u \ast u_{xx} + 70 \, u^{\ast 3} ) \, \pa \nonumber \\
           & & + \frac{1}{64} ( 63 \, u_{xxxxx} + 56 \, u_{xxx} \ast u + 189 \, u_{xx} \ast u_x
               + 231 \, u_x \ast u_{xx} + 35 \, u_x \ast u^{\ast 2} \nonumber \\
           & & + 154 \, u \ast u_{xxx}
               + 70 \, u \ast u_x \ast u + 105 \, u^{\ast 2} \ast u_x )  \; .
\ee
The first three non-trivial equations resulting from (\ref{ncKdVh}) are
\be
 u_{t_3} &=& 2^{-2} \, ( u_{xx} + 3 u^{\ast 2} )_x \\
 u_{t_5} &=& 2^{-4} \, \Big( u_{xxxx} + 5 \, ( u \ast u_{xx} + u_{xx} \ast u ) + 5 \, u_x{}^{\ast 2}
             + 10 \, u^{\ast 3} \Big)_x \\
 u_{t_7} &=& 2^{-6} \, \Big( u_{xxxxxx} + 7 \, (u \ast u_{xxxx} + u_{xxxx} \ast u)
             + 14 \, (u_x \ast u_{xxx} + u_{xxx} \ast u_x) \nonumber \\
         & & + 21 \, u_{xx}{}^{\ast 2} + 7 \, ( 3 \, u^{\ast2} \ast u_{xx}
             + 4 \, u \ast u_{xx} \ast u + 3 \, u_{xx} \ast u^{\ast2})
                          \nonumber \\
         & & + 14 \, ( 2 \, u_x{}^{\ast2} \ast u + u_x \ast u \ast u_x
             + 2 \, u \ast u_x{}^{\ast2}) + 35 \, u^{\ast4} \Big)_x
\ee
starting with the ncKdV equation, and from (\ref{u_theta_odd}) we obtain
\be
   u_{\theta_{1,3}} &=& 2^{-3} \, ( [u_x , u ]_\ast )_x   \label{u_theta13} \\
   u_{\theta_{1,5}} &=& 2^{-5} \, ( [u_{xxx} , u ]_\ast - [u_{xx} , u_x ]_\ast
                        + 5 \, [u_x , u^{\ast 2} ]_\ast )_x
\ee
These equations have already been found in \cite{DMH04hier} by reduction of the ncAKNS
hierarchy. They are recovered from (3.46)-(3.48) and (3.49)-(3.50) in \cite{DMH04hier}
via $u \mapsto -u$, $t_{2n+1} \mapsto 2^{-2n} \, t_{2n+1}$, $\theta_{2n+1} \mapsto
2^{-2n} \, \theta_{1,2n+1}$. Furthermore,
\be
   u_{\theta_{3,5}} &=& 2^{-7} \, \Big( [u_{xxx} , u_{xx} ]_\ast
                        + 3 \, [u_{xxx} , u^{\ast 2} ]_\ast
                        + 6 \, [u_x , u^{\ast 3} ]_\ast \nonumber \\
                    & & + 4 \, ( u \ast u_x \ast u_{xx} - u_{xx} \ast u_x \ast u )
                        + 2 \, ( u_x \ast u \ast u_{xx} - u_{xx} \ast u \ast u_x ) \nonumber \\
                    & & + 12 \, ( u \ast u_x \ast u^{\ast 2} - u^{\ast 2} \ast u_x \ast u )
                        \Big)_x  \; .
\ee

\section*{Appendix F: Dorfman-Fokas recursion operator}
\setcounter{equation}{0}
\setcounter{section}{6}
\addcontentsline{toc}{section}{\numberline{}Appendix F: Dorfman-Fokas recursion operator}
In this appendix we recall some results from \cite{Dorf+Foka92} and draw the relation
with the present work.
In the following, the operator $D_x$ acts as differentiation with respect to $x$
on the ring ${\cal R}[\pa_y]$ generated by $\pa_y$ and ($N \times N$-matrices of)
smooth functions of $x$ and $y$.
In particular, $D_x \pa_y = 0$. $D_x^{-1}$ denotes the formal inverse of $D_x$ (integration).
Furthermore, we introduce the adjoint actions $\mbox{ad}_V W = V \ast W - W \ast V$ and
$\mbox{ad}^+_V W = V \ast W + W \ast V$. Using
\be
   \Psi = D_x^2 - D_x^{-1} \, \mbox{ad}^+_V \, D_x - \mbox{ad}^+_V
          + D_x^{-1} \, \mbox{ad}_V \, D_x^{-1} \, \mbox{ad}_V
\ee
where
\be
    V = - 2 \, u_2 + \pa_y = - 2 \, \phi_x + \pa_y
\ee
one defines recursively operators $S_n$ (acting on ${\cal R}[\pa_y]$) via
\be
   S_0 = \Psi + 2 \, R_{\pa_y}    \qquad
   S_{n+1} = \Psi \, S_n + 4 \, S_n \, R_{\pa_y}  \; .
\ee
Here $R_{\pa_y}$ is the operator of right multiplication by $\pa_y$.
Acting on $1$, the operators $S_n$ produce functions, i.e. elements of
${\cal R}[\pa_y]$ which do not contain $\pa_y$.
In \cite{Dorf+Foka92}, the series of equations
\be
     \phi_{t_{2n+1}} = 4^{-(n+1)} \, S_n 1  \qquad n = 0,1,2, \ldots
     \label{DF}
\ee
has been  named \emph{generalized KP hierarchy}. It corresponds, however,
only to half of the (noncommutative) KP hierarchy, as made manifest by the
notation used in (\ref{DF}). Although the ncKP equation (\ref{ncKPpot})
is easily recovered from the above formulae, the computation of higher
odd ncKP equations turns out to be very time-consuming.
\vskip.1cm

Note that, for example,
$\Phi^{(3)} = D_x^{-1} \mbox{ad}_V \Phi^{(2)} = (D_x^{-1} \mbox{ad}_V)^2 \Phi^{(1)}$
with the $\Phi^{(n)}$, $n=1,2,3$, defined in section~\ref{sec:xncKP}.
Furthermore, we have the following result.
\vskip.2cm

\noindent
{\bf Lemma.}
\be
    \phi_{\theta_{1,n}}
  = {1 \over 2} ( D_x^{-1} \mbox{ad}_V \phi_{t_n} - \phi_{t_{n+1}} )
     \qquad n = 2,3, \ldots  \; .   \label{phi_theta1n_DF}
\ee
\vskip.1cm
\noindent
{\em Proof:} (\ref{eta_integrability}) with $i=2$ and $m=1$ reads
\bez
    \eta^{(n)}_{2,x} = \eta^{(1)}_{2,t_n} - [\phi_x , \phi_{t_n}]_\ast
  = {1 \over 2} \phi_{y t_n} + {1 \over 2} \phi_{xx t_n} - [\phi_x , \phi_{t_n}]_\ast
\eez
using (\ref{eta-coeffs}). By application of the $\omega$-involution
(see section~\ref{sec:recursion}), this leads to
\bez
   \sigma^{(n)}_{2,x}
 = - {1 \over 2} \phi_{y t_n} + {1 \over 2} \phi_{xx t_n} + [\phi_x,\phi_{t_n}]_\ast \; .
\eez
The difference of both equations is
\bez
  ( \eta^{(n)}_2 - \sigma^{(n)}_2 )_x = \phi_{y t_n} - 2 \, [\phi_x , \phi_{t_n}]_\ast
\eez
and thus
\bez
  \eta^{(n)}_2 - \sigma^{(n)}_2 = D^{-1} \mathrm{ad}_V \, \phi_{t_n} \; .
\eez
Inserting this expression in (\ref{phi_theta_1n}), we get (\ref{phi_theta1n_DF}).
\hfill $\blacksquare$
\vskip.2cm

The proof of this Lemma can be easily generalized to obtain a corresponding
expression for $\phi_{\theta_{2,n}}$:
\be
   \phi_{\theta_{2,n}} &=& - \frac{1}{2} \phi_{t_{n+2}}
   + \frac{1}{4} \phi_{t_n xx} + \{ \phi_x , \phi_{t_n} \}_\ast
   - \frac{1}{2} \, D_x^{-1} \{ \phi_{xx} ,  \phi_{t_n} \}_\ast
   + \frac{1}{4} (D_x^{-1} \mbox{ad}_V)^2 \phi_{t_n}   \nonumber \\
  &=& - \frac{1}{2} \phi_{t_{n+2}}
   + \frac{1}{4} \Psi \, \phi_{t_n}
   + \frac{1}{2} \, \{ \pa_y , \phi_{t_n} \}  \; .
\ee


\end{document}